\documentclass[superscriptaddress,twocolumn,showkeys,10pt,pre,aps,nofootinbib]{revtex4-1}

\usepackage{bm}

\usepackage{anysize}
\usepackage[english]{babel}
\usepackage{amsmath,amssymb,latexsym}
\usepackage{verbatim}
\usepackage[utf8]{inputenc}
\usepackage{pst-all}
\usepackage{pstricks}
\usepackage[dvips]{graphicx}
\usepackage{psfrag}
\usepackage{setspace}
\usepackage{fullpage}
\usepackage{subfig}
\usepackage{cancel}
\usepackage{float}
\usepackage{hyperref}
\usepackage{color}
\usepackage{tikz}
\usetikzlibrary{arrows}

\marginsize{1cm}{1cm}{1cm}{1cm}

\newcommand{\suchthat}{\mathrel{\ooalign{$\ni$\cr\kern-1pt$-$\kern-6.5pt$-$}}}

\newpsobject{grilla}{psgrid}{subgriddiv=1,griddots=10,gridlabels=6pt}

\def\({\left(}
\def\){\right)}
\def\[{\left[}
\def\]{\right]}
\def\|l{\left|}
\def\|{\right|}
\def\<{\left <}
\def\>{\right>}
\def\emp{\begin{equation}} 
\def\emps{\begin{equation*}}
\def\fin{\end{equation}} 
\def\fins{\end{equation*}}
\def\eemp{\begin{equation}\begin{aligned}} 
\def\eemps{\begin{equation*}\begin{aligned}}
\def\ffin{\end{aligned}\end{equation}} 
\def\ffins{\end{aligned}\end{equation*}}
\newcommand{\Ttilde}[1]{\tilde{\mathbf{#1}}}
\newcommand{\Ccheck}[1]{\check{\mathbf{#1}}}

\newcommand{\ImT}[4]{\begin{figure}[h!]
\centering
\includegraphics[width=#1\textwidth]{#2}\\
\caption{\small #4}
\label{fig:#3}
\end{figure}}

\newcommand{\ImTR}[5]{\begin{figure}[h!]
\centering
\includegraphics[width=#1\textwidth,angle=#5]{#2}\\
\caption{\small #4}
\label{fig:#3}
\end{figure}}

\begin{document}

\title{Counter-ion density profile around charged cylinders: the strong-coupling needle limit}

\author{Juan Pablo Mallarino}
\author{Gabriel T\'ellez}
\affiliation{Departamento de F\'{\i}sica, Universidad de los Andes - Bogot\'a, Colombia}

\author{Emmanuel Trizac}
\affiliation{Laboratoire de Physique Th\'eorique et Mod\`{e}les
  Statistiques ({\it UMR CNRS 8626}), Universit\'e Paris-Sud, F-91405 Orsay, France}

\date{\today}

\begin{abstract}
Charged rod-like polymers are not able to bind all their neutralizing counter-ions:
a fraction of them evaporates while the others are said to be condensed. 
We study here counter-ion condensation and its ramifications, both numerically
by means of Monte Carlo simulations employing a previously introduced 
powerful logarithmic sampling of radial coordinates, and analytically,
with special emphasis on the strong-coupling regime. We focus on the 
thin rod, or needle limit, 
that is naturally reached under strong coulombic couplings,
where the typical inter-particle spacing $a'$ along the rod is
much larger than its radius $R$. 
This regime is complementary and opposite to the
simpler thick rod case where $a'\ll R$. We show that due account of counter-ion evaporation,
a universal phenomenon in the sense that it occurs in the same clothing 
for both weakly and strongly coupled systems, allows to obtain
excellent agreement between the numerical simulations and the strong-coupling calculations.
\end{abstract}


\maketitle

\section{Introduction}

Some linear biopolymers are intrinsically stiff objects, further rigidified by electric
charges along their backbone. This is the case of double stranded DNA, tubulin, actin, and some viruses.
These macromolecules may be envisioned, to first approximation, as straight charged cylinders, 
attracting neutralizing counter-ions though a logarithmic potential. It was first 
realized by Onsager that this functional form is similar to that of the confinement entropy,
so that not all counter-ions are bound to the polymer 
\cite{manning1969limiting1,*manning1969limiting2,*manning1969limiting3}: a finite fraction only
remains confined in the limit of infinite system size, with no external boundary;
above a certain critical temperature, all counter-ions evaporate. 

The phenomenon of counter-ion condensation/evaporation has been central in a consequential
number of studies since the 1970s, for it is cardinal in a wealth 
of static and dynamic properties of charged polymers 
\cite{manning1969limiting1,*manning1969limiting2,*manning1969limiting3,oosawa1971polyelectrolytes,Rama83,enum1,*enum2,*Shkl99,*enum3,*enum4,*deserno2001osmotic,*enum5,*enum6,naji2005counterions,trizac2006onsager,*TeTr06b}, see e.g.
\cite{Levi02,GrNS02,naji1,BKNN05,Mess09} for more exhaustive references. It is 
governed by the so-called Manning parameter $\xi$, which is the dimensionless line charge 
of the rod: $\xi = q l_B \lambda$, defined from the the valency $q$ of counter-ions, 
the linear charge $\lambda e$, and the Bjerrum length $l_B = e^2/(4\pi \epsilon kT)$
where $\epsilon$ is the dielectric constant of the medium,
$e$ is the elementary charge and $kT$ denotes thermal energy.
Interestingly, the mean-field (Poisson-Boltzmann \cite{Ande06}) scenario of a
complete evaporation of ions for $\xi<1$ in an infinite system, and
of a partial condensation for $\xi>1$ also holds beyond mean-field 
\cite{naji2005counterions,naji1},
when Coulombic couplings are important. Such couplings are conveniently quantified 
by the parameter $\Xi=q^2 l_B \xi/R$ where $R$ is the cylinder radius.
It is defined to match its planar counterpart 
$\Xi = 2 \pi q^3 l_B^2 \sigma$ \cite{BKNN05,trizac1}, where $\sigma e$ is the surface charge of the 
colloid considered, with here $\sigma = \lambda/(2\pi R)$. 

\ImTR{0.37}{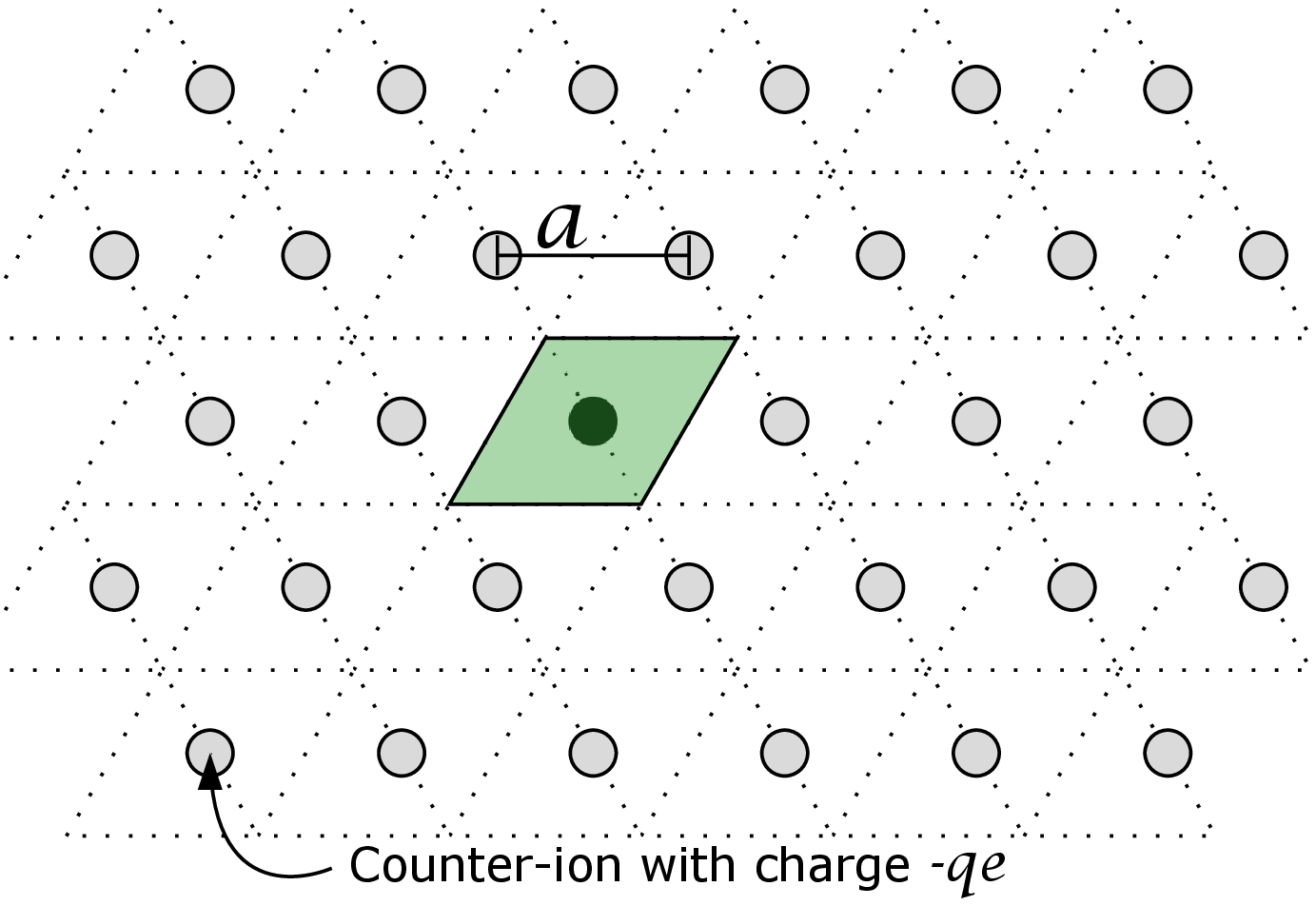}{WignerCrystal}{The Wigner crystal formed for $\Xi\to \infty$ 
at weakly curved
quasi-planar cylinder (thick cylinder case, $a \ll R$).  
The shaded region is the cell area per counter-ion at the surface.
Up to a numerical prefactor, we have $a \propto \sqrt{q/\sigma}$.
At finite but large $\Xi$, such an idealized configuration is met
for $\xi \gg \Xi^{1/2}$.}{0}

While $\xi$ and $\Xi$ both measure the inverse temperature, their
scaling with $1/T$ differ, and their roles in the forthcoming analysis are somewhat
asymmetric. In essence, $\Xi$ governs the distance to mean-field, in the sense that the
Poisson-Boltzmann theory becomes exact for $\Xi\to 0$,
and remains accurate for small $\Xi$. In the present work, we will focus on the opposite
limit of large $\Xi$ values, which defines the strong-coupling regime for which
mean-field is invalid. One may naively think that 
the Manning parameter should also be large for a strong coupling approach
to hold, but we will see that the situation is more subtle, and some quantities
can be obtained for arbitrary
values of $\xi$ provided $\Xi$ is large. It is important though to
clearly discriminate the thin and thick cylinder cases, because they involve 
different mechanisms.
To see this, we start with a cylinder with large radius, where ``large'' means
that $R$ significantly exceeds the typical distance $a$ between ions when 
they are close to their ground state. The situation is locally that 
depicted in Fig. \ref{fig:WignerCrystal} and is mostly governed by the planar
geometry physics studied in \cite{netz2001electrostatistics,KTNB08,trizac1,SaTr11PRE},
up to some curvature corrections that have not been studied so far.
From the electro-neutrality requirement $\sigma a^2 \propto q$, we obtain 
that $\xi \gg \Xi^{1/2}$ for thick rods. 
For the sake of the discussion, we restrict here on ground state considerations, 
and will address thermal effects in detail later on. 
Gradually decreasing $R$ at fixed $\sigma$ (with fixed $\Xi$ and decreasing $\xi$),
one encounters typical configurations
such as that depicted in Fig. \ref{fig:IntermGroundState} where $a$ and $R$ 
are comparable, before reaching the thin or needle case sketched in Fig. \ref{fig:needle},
where $\xi \ll\sqrt\Xi$ ($a\gg R$), and that will be the center of our interest. 
The reason is that a common and experimentally relevant way
to raise Coulombic couplings in a soft matter system is to 
increase the counter-ion valency $q$. Given that $\Xi^{1/2}/\xi \propto \sqrt{q/(\lambda R)}$,
we see that this ultimately leads to $\xi \ll\sqrt\Xi$.
It may be noted that the ground state, reached
e.g. at $T\to 0$, corresponds to both diverging $\Xi$ and $\xi$ parameters,
but the ratio $\Xi^{1/2}/\xi$ is a geometric quantity, independent of the
temperature. In the needle limit, the relevant length scale to measure counter-ion
distances is no longer $a$, defined as a two-dimensional quantity, but its one
dimensional counterpart $a'$ (see Fig. \ref{fig:needle}). We can obtain 
an order of magnitude assuming all ions are condensed onto the rod,
which leads to\footnote{It is customary to introduce the Gouy length $\mu=(2\pi l_B\sigma q)^{-1}$.
We then have $\Xi= l_Bq^2 /\mu$ \cite{moreira2000strong}, and likewise
$\xi = R/\mu$ \cite{naji2005counterions}. Since the interesting regime is for $\xi>1$,
$\mu$ is always the smallest length scale in the problem. In the needle limit
where $\Xi^{1/2}\gg \xi$, we have in general $\mu < R \ll a \ll a'$. 
The relation between $a'$ and $q^2 l_B$ is given by $\xi$,
with roughly $q^2 l_B/a' \simeq \xi$ (taking due account of counter-ion 
condensation, which affects $a'$, we get $q^2 l_B/a' = \xi-1$).}
 $a' \simeq q/\lambda$. This is a lower bound, since the phenomenon
of counter-ion evaporation invariably leads to a lower linear charge than $\lambda e$.

\ImTR{0.2}{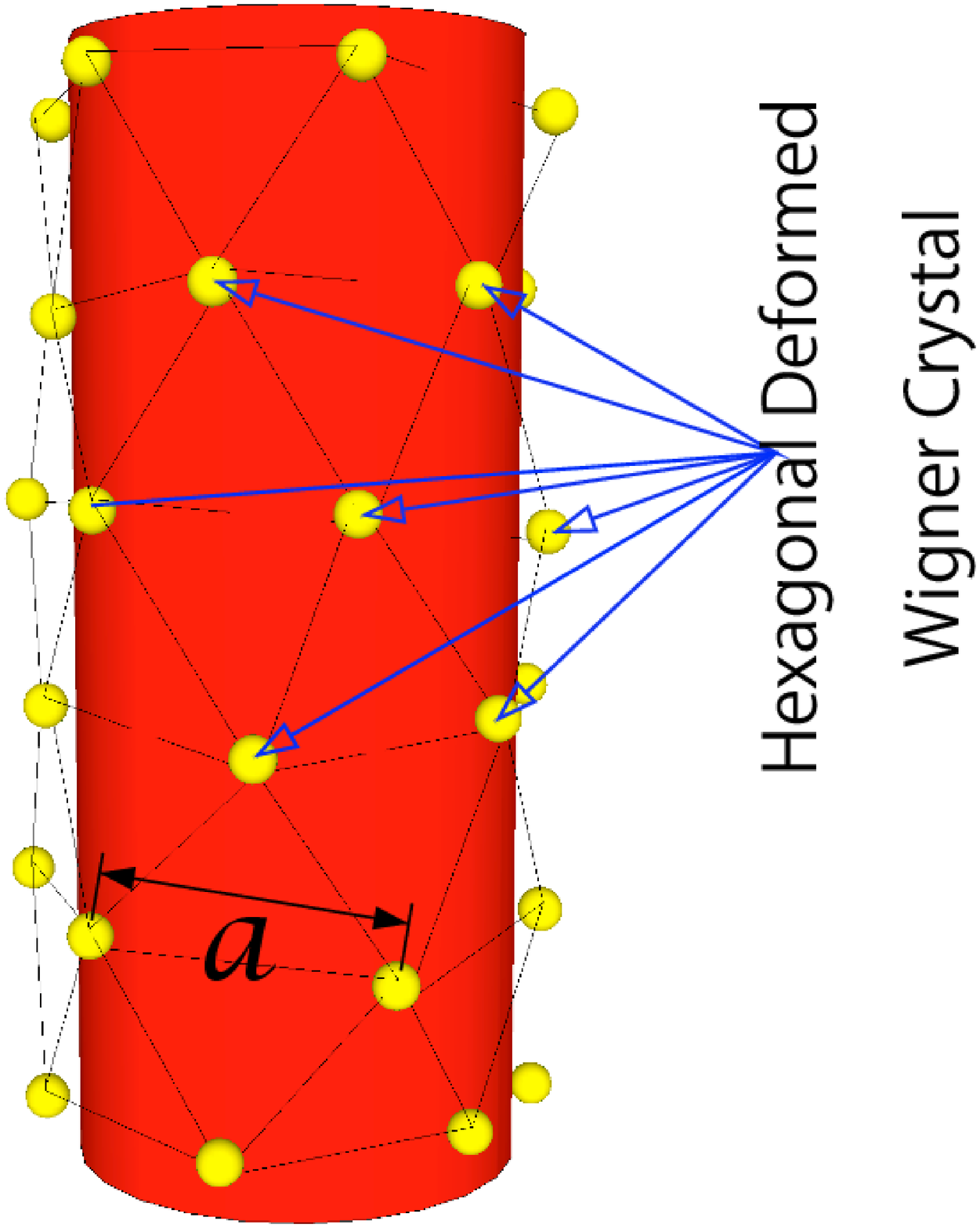}{IntermGroundState}{Artist's view of couter-ions at the charged cylinder 
for large $\Xi$, with $a$ and $R$ of like magnitude. 
Here, $\xi$ is comparable to $\Xi^{1/2}$ }{0}

\ImTR{0.07}{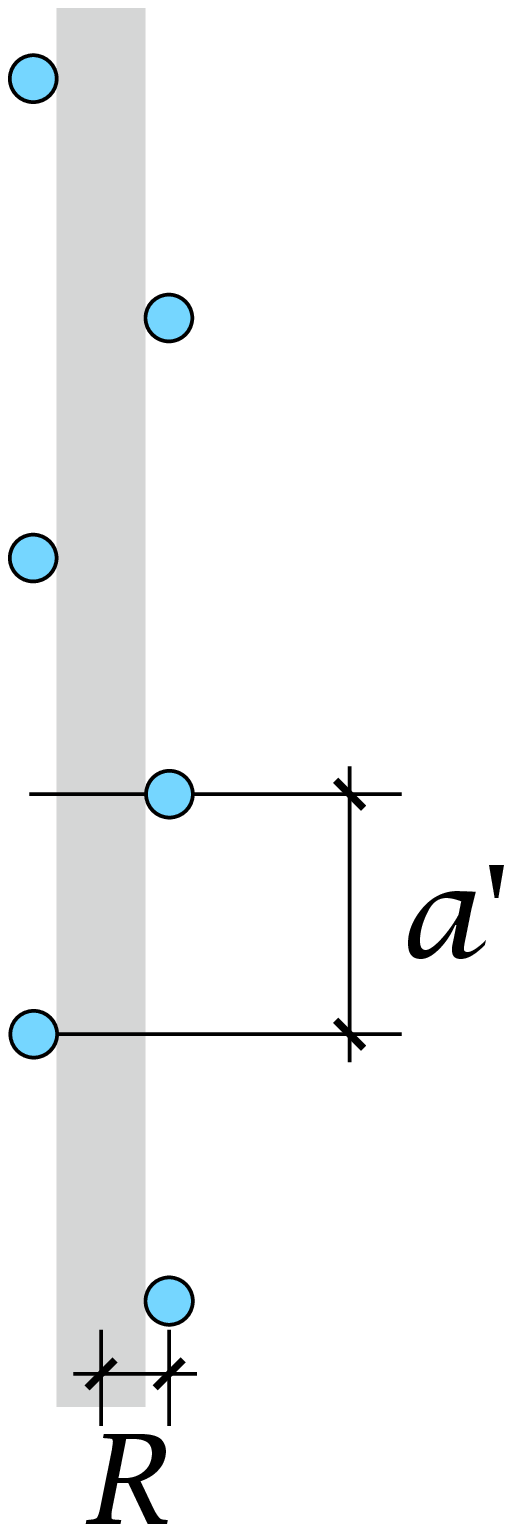}{needle}{Schematic representation of the ground state 
in the {\it needle limit}, that is for $\xi \ll \sqrt\Xi$. 
The distance between charges is denoted
$a'$, and we have $a' \gg a \gg R$.}{0}

As alluded to above, we consider an infinitely long charged cylinder of radius 
$R$, within the primitive cell model (see Fig. \ref{fig:CellModel}): point counter-ions 
with charge $-qe$ are confined in a coaxial larger cylinder of radius $D$.
The ions may equally have a small hard-core, which is immaterial here. We thus
deal with a salt-free system. 
The solvent is accounted for though its uniform dielectric constant $\epsilon$.
No dielectric discontinuity is considered here between the solvent and the 
charged cylinder.
Charged species interact with three-dimensional Coulomb potential,
varying as inverse distance for point particles, with an additional hard-core
term that prevents the ions from entering the charged cylinder. 

\ImTR{0.3}{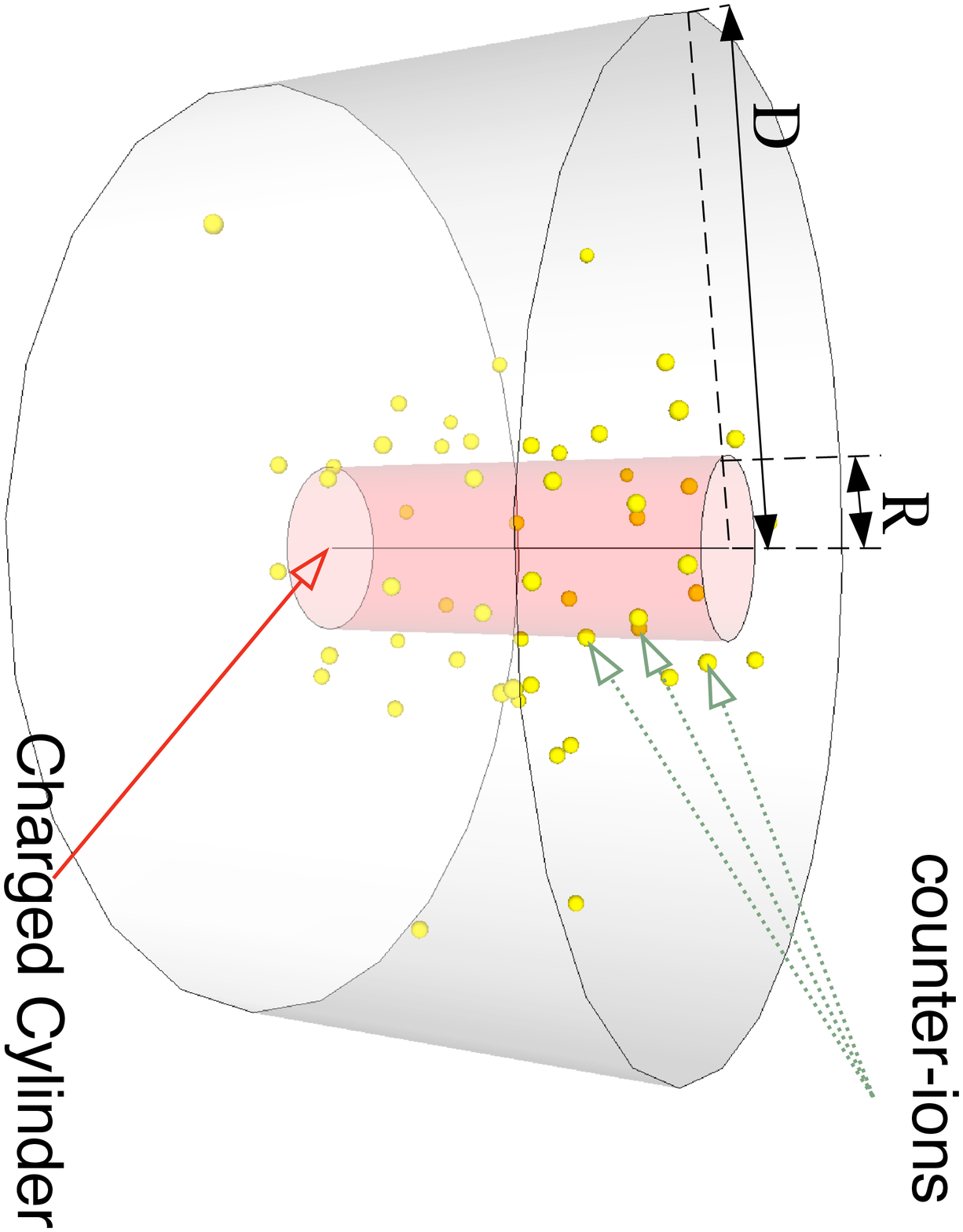}{CellModel}{The 3D cylindrical cell model. The rod is assumed positively charged,
and the counter-ions are therefore negative, with charge $-qe$.}{90}

The outline of the paper is as follows. We first remind some relevant and known results
pertaining to mean-field in section \ref{sec:mf}, 
before presenting in section \ref{sec:sc} our strong coupling (SC) analysis. 
For a given value of Manning parameter $\xi$, we work out in a first step the 
leading order behavior of the density profile when the coupling parameter 
$\Xi \to \infty$.
The present problem in the needle limit is a case where in principle and for the
leading order only,
the virial strong-coupling approach put forward by Netz and collaborators
\cite{moreira2000strong,moreira2001binding,BKNN05,naji2005counterions,naji1,netz2001electrostatistics,KTNB08}
should coincide with its Wigner-SC counterpart worked out in 
\cite{trizac1,SaTr11PRE}. However, due to a different treatment
of counter-ion evaporation, our leading SC expressions (SC-0) will differ 
from previously published ``virial'' results \cite{naji2005counterions,naji1}.
In a second step, we will derive the next correction to SC-0
in the strong-coupling expansion, following the Wigner picture of
\cite{trizac1,SaTr11PRE}, since it has been shown that the virial
approach fails in this task \cite{SaTr11PRE}. Section
\ref{sec:MC} will contain the essentials of the Monte Carlo method used for numerical simulations.
Since the evaporation phenomenon exhibits particularly pronounced finite-size
effects, we will resort to exponentially large system sizes [with typically 
$\log (D/R)$ on the order of a few hundreds], adopting the efficient centrifugal sampling
scheme devised in Ref. \cite{naji1}. The comparison between the analytical predictions 
and simulation data will be provided in sections \ref{sec:results} and \ref{sec:finitesize}.
Some emphasis will be on density profiles, but an order parameter 
for the condensation phenomenon and pair correlation properties will 
also be analyzed (section \ref{sec:results}). While most simulations
have been performed with  system sizes that are large enough and not plagued 
by finite-size effects, the consequences of decreasing the system size 
will be addressed in section \ref{sec:finitesize}.
Finally, section \ref{sec:concl} contains our conclusions.
The technical content of the presentation has been lightened by delegating 
details to appendices.

\section{Mean Field}
\label{sec:mf}

We recall in this section some known results from the mean-field (Poisson-Boltzmann) theory 
\cite{Ande06}. 
The non-linear Poisson-Boltzmann equation admits an analytic solution often attributed to Katchalski 
{\it et. al.} \cite{katchalsky1971poly,fuoss1951potential,lifson1954electrostatic},
but which seems to date back to Liouville \cite{Liou1853}. This solution brings to the fore the importance
of a large lateral extension parameter [$\Delta=\log (D/R)$ with $R$ and $D$ the charged cylinder radius and the outer cylinder cell radius respectively, see Fig.~\ref{fig:CellModel}]. Poisson-Boltzmann equation reads, outside the charged cylinder ($r\geq R$)
\emp
\nabla_{\widetilde{r}}^{2}u(\Ttilde{r})=\widetilde{k}_{D}^{2}e^{u(\Ttilde{r})}
\label{MF}
\fin
where $u(\widetilde{r})=-\beta qe\Phi(r)$ is the dimensionless potential, and $r$ 
the radial distance. Tilde distances are made dimensionless with 
the Gouy-Chapman length $\mu=R/\xi$ ($\widetilde{r}=r/\mu$, $\widetilde{R}=R/\mu=\xi$), and $\widetilde{k}_{D}$ is 
a constant that has no significance before a reference value (a gauge) is chosen 
for the potential. Eq.~(\ref{MF}) is supplemented with the boundary conditions $\partial_{\widetilde{r}} u(\widetilde{D})=0$ (global neutrality of the cell) and $\widetilde{R}(\partial_{\widetilde{r}}u(\widetilde{R}))=-2\xi$ (from Gauss' theorem, 
normal component of the electric field proportional to the surface
charge of the cylinder).
 
The analytic solution depends on the Fuoss critical parameter $\xi_c=\Delta/(1+\Delta)$. Here,
\eemp
u(\widetilde{r})=\begin{cases}
-\log\[\frac{k_{D}^{2}\widetilde{r}^2}{2\alpha^2}\sinh^2\(\alpha\log\frac{\widetilde{r}}{\widetilde{R}}+\coth^{-1}\frac{\xi-1}{\alpha}\)\],&\mbox{if } \xi\leq\xi_c
\\ -\log\[\frac{k_{D}^{2}\widetilde{r}^2}{2\alpha^2}\sin^2\(\alpha\log\frac{\widetilde{r}}{\widetilde{R}}+\cot^{-1}\frac{\xi-1}{\alpha}\)\],&\mbox{if } \xi\geq\xi_c \\
\end{cases}
\ffin
where $\alpha$ is given by the transcendental equations
\eemp
\xi=\begin{cases}
\frac{1-\alpha^2}{1-\alpha\coth(-\alpha\Delta)},&\mbox{if } \xi\leq\xi_c \\
\frac{1+\alpha^2}{1-\alpha\cot(-\alpha\Delta)},&\mbox{if } \xi\geq\xi_c\ .\\
\end{cases}
\label{xi_alpha}
\ffin
The corresponding dimensionless density 
$\widetilde{\rho}=\rho / (2\pi\l_B \sigma^2)$ reads
\eemp
\widetilde{\rho}(\widetilde{r})=\frac{\alpha^2}{\widetilde{r}^2}\times\begin{cases}
\sinh^{-2}\(\alpha\log\frac{\widetilde{r}}{\widetilde{R}}+\coth^{-1}\frac{\xi-1}{\alpha}\),&\mbox{if } \xi\leq\xi_c \\
\sin^{-2}\(\alpha\log\frac{\widetilde{r}}{\widetilde{R}}+\cot^{-1}\frac{\xi-1}{\alpha}\),&\mbox{if } \xi\geq\xi_c \\
\end{cases}
\label{rho_mf}
\ffin
with a normalization condition
\emp
\int_{\widetilde{R}}^{\widetilde{D}}\widetilde{\rho}(\widetilde{r})\widetilde{r}d\widetilde{r}=\xi.
\label{norm_cond}
\fin
A related quantity of interest is the counter-ion integrated charge 
in a cylinder of varying radius, which is, using Gauss' law,
\eemp
\frac{\lambda(\widetilde{r})}{\lambda}&=1+\frac{\widetilde{r}u^\prime(\widetilde{r})}{2\xi}\\
&=1-\frac{1}{\xi}\left[1+\alpha\cot\left(\alpha\ln\frac{r}{R}+\cot^{-1}\frac{\xi-1}{\alpha}\right)\right].
\label{integrated_charge}
\ffin
Electro-neutrality imposes that $\lambda(\widetilde D)=\lambda$
while $\lambda(\widetilde R)= 0$. 

The choice of units ($2\pi l_B \sigma^2$) to measure the density is
of course not essential, but proves convenient in that it will make
contact densities at $r=R$  of order one. In addition, the contact theorem \cite{BlHL79}
imposes that in the limiting case of 
an isolated charged plate, the contact density is strictly fixed to unity: 
$\widetilde\rho=1$. For a given surface charge $\sigma$, the 
planar limit is obtained taking $R\to\infty$, other parameters being
kept constant. It thus corresponds to $\xi\to\infty$ but $\Xi$ fixed, 
a thick cylinder case indeed ($\xi \gg \Xi^{1/2}$). However, 
as far as mean-field is concerned, the difference between thick and thin
cylinders is immaterial, so that we should soon check that $\widetilde \rho(\widetilde R) \to 1$
when $\xi\to \infty$ after having sent the boundary to 
infinity ($\Delta \to \infty$). The SC profiles to follow in section 
\ref{sec:sc} are not endowed with the same property, since the functional 
forms of $\widetilde\rho$ strongly differ in the thick and needle
configurations.

For the most part, the interesting regime is that of $\xi>\xi_c$.
In the large $\Delta = \log(D/R)$ limit, we then have
$\cot(\alpha\Delta)\approx1/(\pi-\alpha\Delta)$ and
\emp
\alpha\approx\frac{\pi}{\Delta+1}\(1-\frac{1}{\xi-1}\frac{1}{\Delta}\),
\fin
which determines the behavior of $\widetilde{\rho}$. From
\eemp
\cot^{-1}\left[\frac{\xi-1}{\alpha}\right]\approx-\alpha(\Delta+1)
\ffin
we get
\emp
\widetilde{\rho}(\widetilde{r})=\frac{1}{\xi^2}\left(\frac{\widetilde{R}}{\widetilde{r}}\right)^2\left[\frac{\frac{\pi}{\Delta+1}}{\sin\left[\frac{\pi}{\Delta+1}\left(\log\frac{\widetilde{r}}{\widetilde{R}}+\frac{1}{\xi-1}\xi_{c}^{-1}\right)\right]}\right]^2,
\fin
which holds for $\xi>1$. Hence, the density in the $\Delta\to\infty$ limit \cite{Rama83}
\eemp
\widetilde{\rho}(\widetilde{r})=\frac{(\xi-1)^2}{\xi^2}&\(\frac{\widetilde{R}}{\widetilde{r}}\)^2\left[1+(\xi-1)\log\frac{\widetilde{r}}{\widetilde{R}}\right]^{-2}.
\label{rho_inf}
\ffin
It appears that for $\xi\leq 1$, $\widetilde{\rho}=0$ at all distances, which signals 
complete evaporation of counter-ions. In other words, a cylinder is only able to 
bind ions if the value of the Manning parameter is higher than unity. 
For $\xi>1$, we further have 
\emp
\int_{\widetilde{R}}^{\infty}\widetilde{\rho}(\widetilde{r})\widetilde{r}d\widetilde{r} \, = \, \xi-1,
\label{new_norm_cond}
\fin
to be compared to (\ref{norm_cond}).
The evaporated fraction of counter-ions is therefore $1/\xi$.
For large although not infinite values of $\Delta$, the phenomenon
remains, although of course, a distance criterion is required 
to differentiate condensed from evaporated ions, since 
normalization (\ref{norm_cond}) always holds. To this end,
a convenient inflection point criterion has often been used 
\cite{deserno2000fraction,qian2000transformed,naji1}
(see also e.g. \cite{TeTr06b} for a related discussion 
with added salt): the integrated charge $\lambda(r)$ plotted as a function 
of $\log r$ shows an inflection point precisely where 
\eemp
\log\frac{r_{c}}{R}=&\Delta\left[1-\frac{\cot^{-1}\alpha}{\alpha\Delta}\right],
\ffin
which corresponds to $\lambda(r_{c})/\lambda = 1-1/\xi$
and renders an effective integrated charge of $\xi-1$. It follows directly that $r_c=R$ at $\xi=1$ and for $\xi>1$
\eemp
\log\frac{r_c}{R}\approx&\frac{\Delta+1}{2}\left[1-\frac{1}{\Delta(\xi-1)}\right].
\label{eq:rc}
\ffin
The distance $r_c$ is often called the Manning radius.

For values of $\xi>1$ ions will be condensed closed to the surface of the cylinder thus creating a cloud of charge with cylindrical symmetry that will screen the potential that other ions farther away perceive. Then, for a sufficiently large distance --beyond $\widetilde{r}_c$-- the integrated charge of the ion cloud and the cylinder is effectively $\xi=1$ (Manning Condensation) and the ions in the outer region will interact with an effective cylinder with charge equivalent to $\xi=1$, which lies at the 
borderline of condensation. From the previous analysis, the fraction of ions condensed within $r_c$ is $f_M$ (Manning condensed fraction of ions), with
\emp
{f}_M=\frac{\xi-1}{\xi},
\label{f_M}
\fin
and the condensed ions will form a cloud of charge $\xi-1$. 
From (\ref{eq:rc}) it appears that the Manning radius $r_c$, for $\Delta$ very large, is close to $\Delta/2$ in logarithmic scale, and the normalization condition 
reads
\emp
\int_{\widetilde{R}}^{\widetilde r_c}\widetilde{\rho}(\widetilde{r})\widetilde{r}d\widetilde{r} \,= \, f_M \xi \, = \, \xi-1,
\label{new_norm_cond_rc}
\fin
to be compared to (\ref{new_norm_cond}) valid for infinite dilution.

On the other hand, to quantify the extension of the electric double-layer, 
it is also instructive to compute the distance at which the integrated charge is half the condensed charge (i.e. $\lambda(r)/\lambda=(\xi-1)/2\xi$). From eq.~(\ref{integrated_charge}),
\eemp
\log\frac{x_{1/2}^{MF}}{R}=\frac{1}{\xi-1+2\frac{\alpha^2}{\xi-1}}\approx\frac{1}{\xi-1}.
\label{eq:x12}
\ffin
This teaches us that the relevant length scale for the extension of the
ionic profile is $R$, a much smaller scale than the Manning radius. 
The above expression is compatible with the known fact 
that in the planar limit, the extension is given by the Gouy length $\mu$.
Indeed, when $R\to \infty$ so that $\xi\to\infty$, we get from 
\eqref{eq:x12} that $x_{1/2}^{MF}-R \propto R/\xi =\mu$.

For future comparison with simulation data, we also precise the behaviour in
the vicinity of the outer cylinder. From eq. \eqref{rho_mf}:
\emp
\xi^2\(\frac{D}{R}\)^2\widetilde{\rho}(r)\approx\left(\frac{D}{r}\right)^2\left[1-\log\frac{r}{D}\right]^{-2}.
\label{rho_edge}
\fin
Conversely, in the vicinity of the charged cylinder when the Manning parameter approaches the critical value - i.e. $\xi\to\xi_c$, we see 
from the transcendental eq. \eqref{xi_alpha} that for $\xi=\xi_{c}^{+}$, $\alpha\to0$ thus yielding
\emp
\xi_{c}^{2}\widetilde{\rho}(\widetilde{r})=\left(\frac{\widetilde{R}}{\widetilde{r}}\right)^2\frac{1}{\left[\Delta+1-\log\frac{\widetilde{r}}{\widetilde{R}}\right]^2},
\label{rho_mf_xi_c}
\fin
which gives a non-vanishing value for the density at contact of $\Delta^{-2}$ for the density. A similar approach to the one performed for $\xi>\xi_c$ renders the same result for $\xi=\xi_{c}^{-}$. Note that 
the value at contact is strongly dependent on the log of the box size.

 
In the following analysis, the results displayed do not depend on the value chosen
for the valency $q$ of counter-ions, which will therefore not be precised,
since it only matters through $\xi \propto q$ and $\Xi \propto q^3$.

\section{Strong Coupling}
\label{sec:sc}

\subsection{Leading order behaviour (SC-0)}
\label{ssec:leading}

In the strong coupling large $\Xi$ limit, at fixed $\xi$, 
the ratio $a'/R \propto \Xi/\xi^2$ becomes large:
this is another way to define the needle limit.
Hence, the typical distance between particles becomes large compared to the radial
distance they explore, and to leading order, the same single particle
picture as in the planar geometry \cite{netz2001electrostatistics,trizac1} does hold \cite{naji1}. 
The ion-ion interactions become sub-dominant compared to the rod-ion term, and
the counter-ion profile is thus given by the exponential of the bare cylinder
logarithmic 
potential. This means that $\rho_0(r) \propto r^{-2\xi}$ \cite{moreira2000strong,naji1}. 
The subscript 0 refers to the dominant order in a large $\Xi$ expansion.
This functional form cannot be normalized for $\xi\leq 1$ --we consider here the infinite
dilution limit--, yet another illustration
of complete evaporation: the omitted prefactor is vanishing for $\xi\leq 1$, so that $\rho_0(r)=0$.
For $\xi>1$, the previous profile is normalizable though, and we have 
\eemp
\widetilde{\rho}_0(r) = \frac{\rho_0(r)}{2\pi l_B\sigma^2}
={f} \, \frac{2(\xi-1)}{\xi}\(\frac{R}{r}\)^{2\xi},
\label{rdf}
\ffin
where we have assumed a fraction of condensed ions $f$, i.e. 
\emp
\int_{\widetilde{R}}^{\infty}\widetilde{\rho}(\widetilde{r})\widetilde{r}d\widetilde{r} \,= \, f \xi.
\label{new_norm_cond_f}
\fin
We have seen in section \ref{sec:mf} that $f=f_M=1-1/\xi$ within mean-field, 
a result that nevertheless holds beyond mean-field \cite{naji1}. 
Taking $f=f_M=1-1/\xi$, we write
\eemp
\widetilde{\rho}_0(r)=\frac{2(\xi-1)^2}{\xi^2}\(\frac{R}{r}\)^{2\xi}.
\label{rdfsc0}
\ffin
This is our leading prediction, denoted  {\bf SC-0} in the remainder,
which turns out to differ from the result derived in \cite{naji2005counterions,naji1}
where the same form as (\ref{rdf}) was considered, but with the choice 
$f=1$
that turns out to be incompatible with evaporation of a non vanishing
fraction of ions. We will refer to the choice $f=1$ as the SC-0 $f=1$ form.
It can be noted here that the typical distance into which the ions are confined 
is given by $R$, and does not depend on $\Xi$. More specifically, 
if we compute the distance corresponding to confinement of 50\% 
of the ions, we get $x_{1/2}^{SC} = R \,2^{(2\xi-2)^{-1}}$,
which exhibits a similar form as its mean-field counterpart (\ref{eq:x12}).
The coupling parameter $\Xi$ is indeed absent, but the expression is compatible 
with the ground state requirement that $x_{1/2}^{SC} \to R$ when temperature
vanishes, because then $\xi\to \infty$. 
For $\xi$ of order 1 but larger than 1 to avoid complete evaporation, 
the relevant confinement scale is $R$. When $\xi$ becomes large, 
we have noted that $x_{1/2}-R \to 0$, and more precisely
$x_{1/2}-R  \simeq R/\xi$, which, again, is the Gouy length 
$\mu \propto (q l_B \sigma)$, setting the confinement range in the planar case.

We note that the contact density following from (\ref{rdfsc0}) reads
$\widetilde \rho_0(R) = 2 (\xi-1)^2/\xi^2$, that is exactly twice the
mean-field contact density found in (\ref{rho_inf}). 
It also appears that it is not possible to recover the planar
limit with its $\widetilde \rho(R)$ constrained to unity by the contact theorem,
since expression (\ref{rdfsc0}) only holds for small values of
$\xi^2/\Xi$ --this is the needle constraint--, 
while the planar limit is met for $\xi\to \infty$ at fixed 
$\Xi$.

\subsection{Correction to leading order (SC-1)}
\label{ssec:sc1}

\ImTR{0.23}{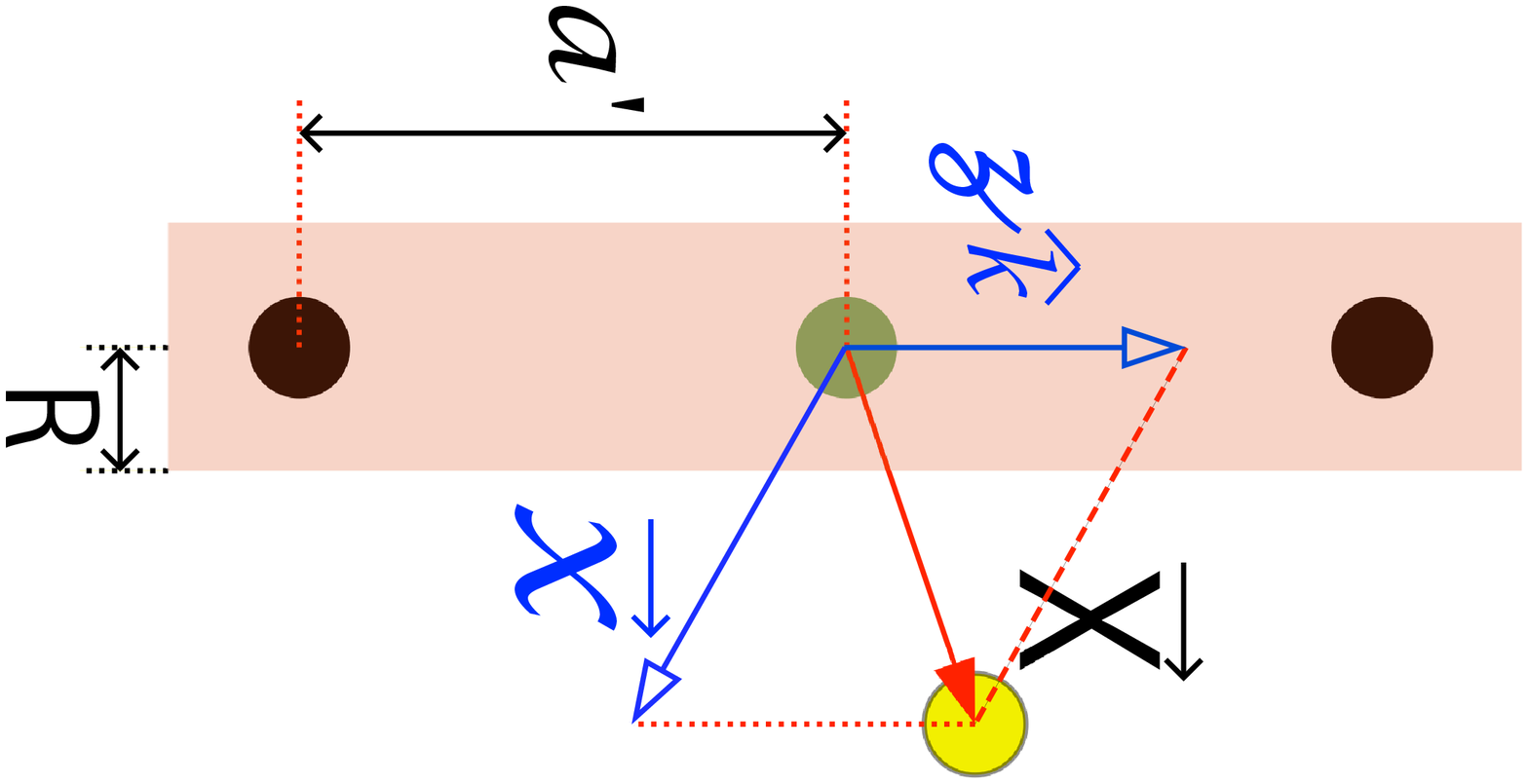}{GroundState}{A $R\to0$ approximation of the ground state for the cylinder system in the needle limit. The length $a'$ should account for ionic evaporation, so that it is defined as $a' = q/(\lambda f)$, with $f=1-1/\xi$. Hence, $R/a' = \xi^2 f /\Xi = \xi(\xi-1)/\Xi$.}{90}

Before comparing our SC-0 prediction to numerical data, we adapt the
method used in Refs. \cite{trizac1,SaTr11PRE} to compute the 
next term in the strong coupling expansion. Such a procedure yields 
fundamentally different results than the virial approach of Refs.
\cite{moreira2000strong,moreira2001binding}, since 
correction terms appear dressed with a different power of $\Xi$.
These corrections have nevertheless not
been worked out at virial SC level in the present cylindrical geometry.

The starting point is to determine the ground state of
the system, and to further consider the relevant 
excitations, those which contribute to the correction to SC-0. 
In other words, we should identify the excitations from particles ground state
that have the smallest energy cost, and we therefore
first expand the inter-particle potential assuming that the particle displacement $\mathbf{X}$ ($\mathbf{X}\colon=\mathbf{x}+z\hat{k}$ $\&$ $X^2=x^2+z^2$, $\mathbf{x}$ being a vector in the plane perpendicular to the $z$ axis) from its lattice position is small, i.e. $\vert\mathbf{X}\vert\ll a^\prime$.
We will assume that ground state positions are given by
$\mathbf{R}_l=a^\prime l\hat{k},\text{ for }l\in\mathbb{Z},
$ 
which corresponds to particles localized onto the $z$ axis, 
see Fig.~\ref{fig:GroundState}. This is consistent with the needle limit
where $R/a' \to 0$ for $\Xi \to \infty$. A more correct ground state is sketched in Fig. 
\ref{fig:needle}, but for computing the desired correction to $\rho_0$,
it is sufficient to distort the true ground state into the simpler form 
shown in Fig. \ref{fig:GroundState}; the same leading correction ensues.

The energy cost for a given configuration of ions reads
(see appendix \ref{U_app} for more details),
\begin{widetext}
\eemp
\beta\delta E=&\xi\sum_{j}\log\(\frac{\widetilde{x}_{j}^{2}}{\widetilde{R}^2}\)+\frac{\xi^3}{\Xi^2}f^3\sum_{j}\left\{-\zeta(3)\widetilde{x}_{j}^{2}+\frac{1}{2}\sum_{l\neq j}\frac{\Ttilde{x}_{l}\cdot\Ttilde{x}_{j}}{\vert j-l\vert^3}\right\}+\xi f\sum_{j}\left\{2\zeta(3)\frac{{z}_{j}^{2}}{a'^2}-{z}_{j}\sum_{l\neq j}\frac{{z}_{l}}{a'^2\,\vert j-l\vert^3}\right\},\\
\label{betaE}
\ffin
\end{widetext}
where $\zeta$ is the Riemann Zeta Function ($\zeta(3) \simeq 1.202$). 
In all formulas and provided that the system size is big enough, 
we consider $f=1-1/\xi$. We however leave the fraction $f$ apparent,
for it becomes a non trivial function of $\xi$, coupling parameter, and confinement
when finite-size effects do matter (see section \ref{sec:finitesize}).

The form of Eq. (\ref{betaE}) calls for some comments, before its use in the 
Wigner strong-coupling machinery \cite{trizac1,SaTr11PRE}. 
When considering separately the displacements
of particles along the rod (variables $z$), or perpendicularly
(variables ${\bf x}$), a dual localization phenomenon appears.
The coupling parameter $\Xi$ governs the localization of ions onto the rod,
which was already clear from the relation $R/a' \propto \xi^2f/\Xi < \xi^2/\Xi$.
It is not coupled to the $z$ degree of freedom, so that even at very large
$\Xi$, the essentially one dimensional system of condensed counter-ions
may be fluid. It is then the Manning parameter $\xi$ that governs
crystallization along the rod direction, and is thus the parameter 
coupled to the $z$ degree of freedom in (\ref{betaE}). Another way to recover
this conclusion is to compute the coupling (plasma) parameter corresponding
to ions on a line, with inter-particle distance $a'$: we get $q^2 l_B / a'$,
which is equal to $\xi-1$. We therefore expect a one dimensional
transition for large values of $\xi$, at large $\Xi$ such that
the needle scenario holds ($\Xi \gg \xi^2$).
In all this discussion, it is implicitly understood that the ions 
are typically confined, radially wise, in a sheath of extension $R$
around the cylinder. This is indeed the case, in the SC regime as well 
as within mean-field, see the discussion in section \ref{ssec:leading}.

Considering the Boltzmann weight constructed from the energy (\ref{betaE}),
we fix one tagged particle at a given position $\mathbf{x}_0$, and integrate over the
remaining particles, in the spirit of the procedure worked out in Refs. 
\cite{trizac1,SaTr11PRE}:
$\rho(\mathbf{x_0})=C\langle\delta(\mathbf{x}-\mathbf{x}_0)\rangle$, with $C$ a normalization constant. 
After some algebra detailed in appendix \ref{U_app}, 
and under the proper normalization condition (eq.~\ref{new_norm_cond}),
we arrive at
\eemp
\widetilde{\rho}_{1}(r)=& 2f\frac{(\xi-1)}{\xi}\(\frac{R}{r}\)^{2\xi}\times\\
&\left\{1+\zeta(3)\frac{\xi^5}{\Xi^2}{f}^3\left[\left(\frac{r}{R}\right)^{2}-\frac{\xi-1}{\xi-2}\right]\right\},\\
\label{rdf_corr}
\ffin
subsequently referred to as {\bf SC-1}. 
We will discuss later the limit of validity of the above expansion. 
It should be emphasized that the profile (\ref{rdf_corr}) is an expansion
in $r$, which does not hold up to $\infty$. This poses a problem
for normalization, since the neglected higher order terms become
prevalent for $\xi<2$, and explain why (\ref{rdf_corr}) can only be normalized
for $\xi>2$. This leads to the conclusion that while the zeroth order
term $\rho_0$ may give a reasonable profile for small values of $\xi$ (to be precised 
in section \ref{sec:results}), the correction $\rho_1$ is deficient 
for $\xi<2$.

\subsection{A single particle variant}

There is an alternate semi-numerical treatment to the strong-coupling 
problem. Within the range of high values of $\xi$ and $\Xi$, such that $\xi^2f/\Xi\ll1$, 
we fix the ions at their ground state positions, and compute the energy cost if one particle
(and only one) is shifted perpendicularly from this structure, consistent with the single particle picture in the strong coupling regime. The density profile is then
\emp
\widetilde{\rho}({r}) \propto\left(\frac{{R}}{{r}}\right)^{2\xi}
\exp\left[-2\xi f\sum_{j>0}\(\frac{1}{\sqrt{j^2+\(\widetilde{r}\frac{\xi f}{\Xi}\)^2}}-\frac{1}{j}\)\right],
\label{rdf_numeric}
\fin
up to a normalization constant.
The series involved in the calculation has no known closed form. For large distances,
\emp
\sum_{j>0}\(\frac{1}{\sqrt{j^2+x^2}}-\frac{1}{j}\)\underset{x\to\infty}{\approx}-\log{x}+\text{const.}
\fin
meaning that the large distance behaviour is $\rho(r) \propto 1/r^2$, as is the case
within mean-field. The profile (\ref{rdf_numeric}) is, therefore, not normalizable 
when $\Delta \to \infty$.
There is however a large range of upper cutoff distances
where the resulting normalized expression (\ref{rdf_numeric}) is invariant close to
the charged rod, so that the normalization problem can be in practice
easily circumvented. Eq. (\ref{rdf_numeric}) can be viewed as an improved version of SC-0, and will be referred to as {\bf SC-0*}. In particular, it reproduces the $r^{-2\xi}$ behaviour in the
vicinity of the charged rod.

\section{Monte-Carlo Simulations}
\label{sec:MC}

For the numerical computations, we will adapt Monte-Carlo sampling to the cell geometry, 
and impose periodic boundary conditions along the main axis of the cylinder ($z$),
taking due account of the long range of Coulomb potential. 
In the $xy$ plane, the counter-ions are confined between the cylinder ($R$) and the outer shell ($D$). There has been extensive reports on numerical implementations for the calculation of the potential for periodic boundary conditions under a number of geometries \cite{mazars2001lekner,mazars2005lekner,deserno1998mesh,arnold2005mmm1d,limbach2006espresso}. For the 1D periodic case, the Lekner-Sperb sums \cite{deserno1998mesh} have been the standard method to account for all the electrostatic contributions. The evaluation of such sums is numerically expensive, 
and requires in particular the calculation of an important number of terms for short distances.
Here, we introduce a novel analytic formulation deduced from the Poisson-Jacobi transformation, as proposed in \cite{mazars2010ewald}.  It is simple to implement and is free of divergences in all ranges of  interparticle distances. The resulting energy is based on the Ewald separation of the potential in a term that converges quickly in real space and another that converges quickly in Fourier space. Details are presented in appendices \ref{1D-Ewald} and \ref{SF}. 

The potential energy of the system can be expanded as
\eemp
U=U_{R}+U_{F}+U_{C}+U_{S}.
\ffin

Each of the terms are written in terms of two conveniently defined variables, $\rho_{ij}$ the distance between the particles' positions projected to the plane perpendicular to the $z$ axis and $z_{ij}=z_i-z_j$; here $L_z$ is the length of the box along the the periodic direction which is naturally chosen as the cylinder axis $z$.

\footnotesize
\eemp
U_{R}=&\frac{1}{4\pi\epsilon\epsilon_0}\sum_{i=1}^{N-1}\sum_{j=i+1}^{N}q_iq_j\left[\sum_{n}\frac{\text{Erfc}\(\alpha\(\rho_{ij}^{2}+(z_{ij}+L_zn)^2\)^{\frac{1}{2}}\)}{\(\rho_{ij}^{2}+(z_{ij}+L_zn)^2\)^{\frac{1}{2}}}\right.\\
&\left.+\frac{1}{L_z}\begin{cases}
0,&\rho_{ij}=0 \\
-\gamma-\log\(\alpha^2\rho_{ij}^{2}\)-\text{E}_1\(\alpha^2\rho_{ij}^{2}\),&\rho_{ij}>0 \\
\end{cases}\right],
\ffin
\normalsize
where $q_i$ is the charge of particle $i$, $\alpha>0$ is a parameter chosen for convergence, $\gamma$ the Euler-Mascheroni constant, $\text{Erfc}(x)$ the complementary error function and $\text{E}_1(x)$ is the exponential integral as defined in Appendix \ref{SF}.

\footnotesize
\eemp
U_{F}=\frac{1}{2\pi\epsilon\epsilon_0L_z}\sum_{i=1}^{N-1}\sum_{j=i+1}^{N}q_iq_j\sum_{k>0}\mathbf{K}_{0}\(\frac{k^2}{4\alpha^2},\alpha^2\rho_{ij}^{2}\)\cos(k\cdot z_{ij}),
\ffin
\normalsize
where $k=2\pi n /L_z$ for $n\in\mathbb{Z}$ and $\mathbf{K}_0(x,y)$ is the incomplete Bessel function. Further references for the evaluation of this function can be found in \cite{Harris2008,Harris2009}.
\eemp
U_{C}=&-\frac{\lambda}{2\pi\epsilon\epsilon_0}\sum_{i=1}^{N}q_i\log\(\frac{\rho_{i}}{R}\),
\ffin
with $\lambda$ the linear charge density of the cylinder ($\sigma=\frac{\lambda}{2\pi R}$). Notice that the cylinder is located at the origin of coordinates.
\footnotesize
\eemp
U_{S}=&\frac{1}{4\pi\epsilon\epsilon_0}\(\sum_{n>0}\frac{\text{Erfc}(\alpha L_zn)}{L_zn}+\frac{1}{L_z}\sum_{k>0}\text{E}_1\(\frac{k^2}{4\alpha^2}\)-\frac{\alpha}{\sqrt{\pi}}\)\times\\
&\(\sum_{i=1}^{N}q_{i}^{2}+\lambda^2L_{z}^2\)+\frac{\lambda^2L_z}{4\pi\epsilon\epsilon_0}(\gamma+\log(\alpha^2R^2)).
\ffin
\normalsize
The previous expression for the energy is free of divergences for any value of $\rho_{ij}=0$ and since the particles are bounded to the cell, we have $0<\log(\rho_{i}/R)<\Delta$. All simulations were taken with a number of particles that ranged between $300$ and $1000$, and ran typically over $10^7$ steps.

In the problem under scrutiny, finite-size effects are important and logarithmic in $D/R$ \cite{naji1},
which requires very large system sizes and consequently precludes standard sampling methods. To circumvent this difficulty, we use the centrifugal sampling technique \cite{naji1} which consists in writing the partition function with more convenient log variables, $Y=\log(r/R)$. 
Then, the partition function 
\eemp
\mathcal{Z}=C_0\int_{V^N}d\mathbf{r}^Ndz^N\exp\(-\beta U(\mathbf{r}^N,z^N)\)
\ffin
transforms to
\eemp
\mathcal{Z}^\prime=C_{0}^{\prime}\int_{{V^\prime}^N}d\theta dY^Ndz^N\exp\(-\beta U(\mathbf{r}^N,z^N)+2\sum_iY_i\),
\ffin
which redefines the energy into
\eemp
U^\prime(Y^N,\theta^N,z^N)=U(\mathbf{r}^N,z^N)-\frac{2}{\beta}\sum_iY_i.
\ffin
The calculation of the inter-particle potential energy requires to know the Cartesian coordinates of the particles; hence, the transformation of coordinates has to be performed each time a particle moves. For the latter choice of variables we can choose a Monte-Carlo step size of $(\Delta Y,\Delta\theta,\Delta z)$ such that for short distances ($Y\approx0$), a displacement of $\Delta r$ is at most of a Gouy length. Equivalently,
\eemp
\Delta Y=\log\left[\frac{R+\mu}{R}\right]\approx\frac{1}{\xi}.
\ffin

For the sake of efficient equilibration, it proves useful to distinguish 
between two kinds of ions in our system.
Bounded ions are constrained to a shell measured in $R$ units as discussed in sections 
\ref{sec:mf} and \ref{ssec:leading}. Their unbounded counter-parts cover a region
beyond the 
Manning radius ($r_c$)
where the average inter-particle spacing is much larger than the electrostatic 
correlation length $l_b$, hence forming a weakly coupled gas. 
From mean field (eq. \eqref{eq:rc}), we expect the 
location of the Manning radius near $\Delta/2$ in log units. 
This question will be further explored when presenting our results.
As a 
part of the equilibration process, particles have to be exchanged between the 
two populations. To this end, 
we propose to move one particle 
from $Y$ to $Y^\prime=\Delta-Y$.
Performing twice such a move will return a particle
to its original position. Therefore, detailed balance is preserved by
choosing a fixed probability $p$ ($p\sim10^{-4}$) to select this type
of particle exchange move over a regular one.  The attempt is then accepted
employing the usual Metropolis criterion. Such an approach guarantees
proper equilibration independently of any chosen initial condition for
the counter-ions. This is particular important for the study of finite
size systems at large coupling, as algorithms with only standard moves
(including the original centrifugal sampling method) will
not sample correctly the configuration space.

\section{Profiles, correlations and order parameter}
\label{sec:results}

We present in this section the bulk of our results. A large system has been simulated, in order to
get rid of finite-size effects, that shall be studied separately
but in a more cursory fashion in section \ref{sec:finitesize}.
We start by validating our simulation procedure against 
known results. To this end, two features can be used.
We can first check that for small $\Xi$, the mean-field (MF) expressions
are recovered, and also that at arbitrary $\Xi$, the condensation
scenario coincides with the MF one \cite{naji1}.

\subsection{Counter-ion condensation / evaporation}

\psfrag{AX}{$\log(r/R)$}
\psfrag{BY}{$\lambda(r)/\lambda$}
\psfrag{AX2}{$$}
\psfrag{BY2}{$$}
\ImT{0.4}{art2_charge}{CumulativeDensity}{Monte Carlo measured cumulative density of particles for $\Delta=300$ as a function of the logarithmic distance $\log(r/R)$, for $\Xi=10^{-1},10^2,10^5$ and 
$\xi=1,2,3,4$. The crosses drawn close to ($\Delta/2,f=1-1/\xi$) 
are the expected locations of the inflection point, at the Manning
radius $r_c$ given by (\ref{eq:rc}). The mean-field prediction (\ref{integrated_charge}) superimposes to the $\Xi=0.1$ results.}

We begin by the condensation phenomenon. For $\xi<1$, all ions dilute away 
when $\Delta = \log(D/R)$ increases. For $\xi>1$, a fraction $f$ of ions
remain condensed in the vicinity of the charged rod. This can be seen
in Fig. \ref{fig:CumulativeDensity}, which shows the integrated line charge
$\lambda(r)$ in a cylinder of varying radius $r$, as introduced in the mean-field
section \ref{sec:mf}. By definition, $\lambda(R)=0$ while electro-neutrality requires
$\lambda(D)=\lambda$.
The ionic atmosphere is more bound to the 
rod as the coupling parameter $\Xi$ increases, and the profile then strongly
departs from MF. Further from the rod, ionic correlations decrease as a consequence 
of the lower ionic density,
to such an extent that the tail of the ionic profile is described
by mean-field. This is why in the right hand side of the figure,
the different $\Xi$-curves collapse, and coincide with MF form.
The inflection point property, that 
is clearly visible, hence takes place at a point that is $\Xi$
independent, for which mean-field results apply \cite{naji1}. 
This provides the rationale for the two-fluid picture
(bound ions before the inflection point, unbound beyond)
that is often used for polyelectrolyte, but that is quite
specific to the salt-free case \cite{TeTr06b}. Note also that to observe the
MF condensed fraction $f=f_M=1-1/\xi$ as in Fig. \ref{fig:CumulativeDensity},
exponentially large box sizes are required. 
We come back to this point in section \ref{sec:finitesize}.
Our results reproduce previously reported data \cite{naji2005counterions,naji1}.
For instance, extracting the inflection-point location from plots
such as Fig. \ref{fig:CumulativeDensity}, we always obtain 
a condensed fraction that is extremely close to $f_M=1-1/\xi$,
see Fig. \ref{fig:Condensedfractionofions}. Figures \ref{fig:CumulativeDensity}
and  \ref{fig:Condensedfractionofions} justify the normalization choice
made in section \ref{sec:sc}, that led to Eqs. (\ref{rdfsc0}) and (\ref{rdf_corr}).
Indeed, the strong-coupling profiles are meant to describe the ionic
atmosphere in the vicinity of the charged rod, an atmosphere that is deprived,
over an exponentially large distance range,
from the ions that lie in the vicinity of the confining border. 
This results in the plateau of Fig. \ref{fig:CumulativeDensity},
at a value that does not correspond to full neutrality.
Note that the term ``vicinity'' here should be taken in the broad
sense, since it can be seen in Fig. \ref{fig:CumulativeDensity} that 
$\lambda(r)$ changes to reach full neutrality $\lambda(D)/\lambda=1$
in the range where $260<\log(r/R)<300$, so that $r$ changes by a factor 
$e^{40}\simeq 10^{17}$.

\ImT{0.4}{art5_f}{Condensedfractionofions}{Condensed fraction of ions as a function of the Manning parameter $\xi$ for values of $\Xi$ in mean field and in the strong coupling 
regime ($\Delta=300$).}

\subsection{Density profiles}

\psfrag{AX}{$r/R$}
\psfrag{BY}{$\widetilde{\rho}(r)$}
\ImT{0.4}{art3_density}{RadialDensity2}{Radial density profile for different values of $\Xi$ and 
fixed $\xi=4$ for $\Delta=\log(D/R)=300$. The symbols show the Monte Carlo results.
The lines are for the the mean field (MF) result and our SC-0 (strong coupling to leading order) analytic solution (\ref{rdfsc0}). The SC-0$f=1$ prescription
of Ref. \cite{naji2005counterions,naji1} is shown by the upper dotted curve. The reduced density is defined 
as $\widetilde{\rho}=\rho / (2\pi\l_B \sigma^2)$.}

Having validated our normalization procedure from integrated profiles, we turn with Fig. 
\ref{fig:RadialDensity2} to a more
precise analysis of the profiles themselves close to the charged cylinder.
As expected, the small $\Xi$ results coincide with their mean-field
form, see the $\Xi=0.1$ curve. On the other hand, when $\Xi$ is large enough 
and exceeds $10^2$, all profiles collapse onto the SC-0 prediction, 
Eq. (\ref{rdfsc0}). As anticipated, the mean-field contact value 
[$\widetilde \rho(R)=9/16 \simeq 0.56$ for $\xi=4$] is half its strong-coupling
counterpart. A more thorough analysis of the contact density will be
presented in section \ref{ssec:close}, 
in conjunction with the study of the corrections to SC-0 and the test 
of the SC-1 formulation. 

The algebraic form of SC-0 is better appreciated in the logarithmic plot 
of Fig. \ref{fig:xi4_log}, which shows that the $\Xi=10^2$ points tend to 
depart slightly from SC-0 for $r>1.6R$, while those for $\Xi=10^3$
are in excellent agreement with the prediction (\ref{rdfsc0}). 
As explained above, Eq. (\ref{rdfsc0}) is a double expansion, first in large 
$\Xi$, and second in small distances. It therefore does not hold up to arbitrary large 
$r$, and we will comment further the large distance behavior in section 
\ref{ssec:universalMF}.

\ImT{0.4}{art3_xi4_log}{xi4_log}{Log-log plot of the radial density profile for different values of $\Xi$ with fixed $\xi=4$. The symbols display the Monte Carlo results.
The dashed line indicates the mean field (MF) prediction coinciding with the
$\Xi=10^{-1}$ Monte Carlo data. The continuous curve is for the SC-0 strong-coupling to leading order analytic solution 
(\ref{rdfsc0}) which superimposes to the $\Xi=10^3$ data 
and partially with the $\Xi=10^2$ points, and finally 
the dotted curve is for the SC-0$f=1$ formula of \cite{naji2005counterions,naji1}.}

Consistently with the integrated charge plateau 
of Fig. \ref{fig:CumulativeDensity}, the $f=1$ normalization of Ref. \cite{naji2005counterions,naji1}
does not fit with the numerical data, see the upper dotted 
curve in Fig. \ref{fig:RadialDensity2}. This phenomenon is all the more
pronounced as the Manning parameter is low, see Figs. \ref{fig:xi1.2}
and \ref{fig:xi1.4}. On the other hand, the SC-0 Eq. (\ref{rdfsc0}) for $f=1-1/\xi$  provides
a reasonable profile when $\Xi$ is large enough, and $\xi$ not too close
to unity: the agreement in Fig. \ref{fig:xi1.4} is correct, and better than in Fig.
\ref{fig:xi1.2}. This was expected, since our SC approach is an expansion 
in the vicinity of the ground state of the system, and therefore better 
when, in addition to $\Xi$, $\xi$ is large enough. It even comes 
as a surprise that we can get semi-quantitative agreement for such 
low values as $\xi=1.2$, and good agreement for $\xi=1.4$. 

\psfrag{AX}{$r/R$}
\psfrag{BY}{$\widetilde{\rho}(r)$}
\ImT{0.4}{art3_xi1.2.eps}{xi1.2}{
Same as Fig. \ref{fig:RadialDensity2}, where the SC-0$f=1$ prediction
of Ref. \cite{naji2005counterions,naji1} is shown by the dotted curve in the inset}

\ImT{0.4}{art3_xi1.4.eps}{xi1.4}{ Same as Fig. \ref{fig:xi1.2}, for $\xi=1.4$.}

\subsection{Universal crossover to mean-field at large distances}
\label{ssec:universalMF}

The decay of the ionic profile with distance $r$ has led various authors to
surmise that mean-field should hold at large enough distances 
\cite{Shkl99,ChWe06,DSDL09}. Indeed, one may define a local 
coupling parameter $\Xi(r)$ from the ratio of Bjerrum length
to the typical distance $\bar d$ between counter-ions at a given distance $r$.
With $\rho(r) \bar d^3$ of order unity, we expect $\Xi(z) \propto \rho(z)^{1/3}$,
which should bring the profile into the mean-field region $\Xi(z) \ll 1$
when $z\to\infty$. 
We emphasize that explicit checks of this expectation are in general difficult to 
perform, due to the fact that the distances one should be able to probe
can be very large. Our present study is nevertheless particularly
well suited for investigating such an effect, due to the large systems
considered.

We start by considering infinite system sizes $\Delta\to\infty$. 
We note here that if mean-field holds at large $r$, then, the profiles 
should not only become independent of $\Xi$, but also on $\xi$
(attention should be paid here to the difference between
the normalized profile $\widetilde \rho$ and the original one 
$\rho \propto l_B \sigma^2 \widetilde \rho $). This is a property
of the relation (\ref{rho_inf}) which gives asymptotically that 
$\widetilde\rho(r) \propto \xi^{-2} (R/r)^{2} / \log^2(r/R)$.
Going back to the initial profile,
we get $\rho(r) \propto l_B^{-1} r^{-2}  / \log^2(r/R)$, for all values of $\xi$.
This universality is illustrated in Fig. \ref{fig:crossMF}, where the
curves for different $\xi$ and for large couplings asymptotically coincide with 
the MF expression. The figure also 
highlights the fact that the distances needed to evidence
the MF form are quite large, $r/R>e^{10}\simeq 2\times10^4$.
The figure corresponds to $\Delta=\log(D/R)=300$, so that for
the range of distances displayed, the behavior is very close 
to its $\Delta \to \infty $ limit. A single mean-field curve appears
in Fig. \ref{fig:crossMF}, since for the range of distances shown,
the MF solution for the different $\xi$ values differ only for very small 
$r$.

\psfrag{AX}{$\log(r/R)$}
\psfrag{BY}{$\xi^2 \,  \widetilde{\rho}(r) \, / q^2$}
\ImT{0.4}{art20a}{crossMF}{Plot of $\xi^2 \widetilde \rho / q^2 = 2 \pi l_B R^2 \rho$
as a function of radial distance, for $\Delta=300$. As in other figures, the 
symbols are for the Monte Carlo data. The mean-field prediction, which holds for the three values
of $\xi$, has been added (line). }

Figure \ref{fig:crossMF} revealed that the mean-field tail is visible for small
densities / large distances only. On closer inspection, it appears that
the departure from the SC behavior which holds
at small $r$ is quite sharp, as shown in Fig. \ref{fig:crossover}: the algebraic
profile in $r^{-2\xi}$ holds up to $r/R \simeq 1.4$. It abruptly evolves into
another form for larger distances, a form that is still far from the MF expression (upper dotted
curve) but  closer to the critical $\xi=1$ mean-field curve (which of course
is fully compatible with the MC results at the low $\Xi=0.1$). The latter 
remark provides an approximate means to compute the crossover point $r_c$ where
SC-0 ceases to hold: we simply equate the SC-0 (eq. \eqref{rdfsc0}) 
and MF-$\xi=\xi_c$ density value at contact (eq. \eqref{rho_mf_xi_c}) forms to get
\eemp
\log x_c=\frac{1}{\xi}\log\left[\sqrt{2}f_M\Delta\right],
\ffin
where $x_c = r_c/R$. We learn here that the dominant form of the crossover point 
behaves as
\eemp
x_c\propto\Delta^{\frac{1}{\xi}}.
\ffin
The dependence on the 
Manning parameter agrees qualitatively with Fig. 
\ref{fig:crossMF} where $x_c$ decreases upon increasing $\xi$.
One interesting trait of finite $\Delta$ on the distribution of ions is that at 
the critical Manning transition parameter ($\xi=\xi_c$) the value of the 
contact density at the surface of the cylinder is not zero but $\Delta^{-2}$,
which can be a small quantity,  thus, 
providing the order of magnitude of the density that must be reached before 
a mean-field like behavior can emerge.

\psfrag{AX}{$r/R$}
\psfrag{BY}{$\xi^2\widetilde{\rho}(r)$}
\ImT{0.4}{art19_crossover}{crossover}{The radial density for different values of 
$\Xi$, $\xi$ and $\Delta=300$. The dashed lines indicate respectively the mean 
field (MF) profile for $\xi=1$ and the dotted curve for $\xi=25$. The strong 
coupling to leading order  analytic solution  (SC-0) is presented in the solid 
curve. The right arrow indicates the crossover location ($x_c\approx1.38$) and the up arrow indicates the location of a Gouy length from the surface of the cylinder.}

A complementary means to illustrate the universality of MF behaviour at large 
distances is provided in Fig. \ref{fig:DensityEdge},
which is explicitly governed by finite size effects. 
We know from section \ref{sec:mf} that the density in the vicinity of
the confining cylinder at $r=D$ is given by Eq. (\ref{rho_edge}). 
The data collapse displayed in Fig. \ref{fig:DensityEdge} is remarkable,
and shows that the $\xi$ and $\Xi$ independent mean-field physics is at work 
in the tail of the profile.

\psfrag{AX}{$\log\left(r/R\right)$}
\psfrag{BY}{$\xi^2(D/R)^2 \, \widetilde{\rho}(r)$}
\ImT{0.4}{art13_rho_SC}{DensityEdge}{Profile near the edge of the box,
for  $\xi=2$ or 3, $\Delta=300$, and values of $\Xi$ in all regimes. 
The solid curve corresponds to the mean-field Eq.~(\ref{rho_edge}).}

\subsection{Order parameter for the evaporation transition}

The mean inverse distance is a parameter that can conveniently be used to see the transition from the condensed to the de-condensed phase for which, a priori, we expect a critical change around $\xi=1$.
We therefore define the order parameter $S_1$ as $N^{-1}\sum_{i=1}^N \overline{\widetilde r_i^{-1}}$,
where the overline refers to the Monte Carlo time average at equilibrium. In other words, 
we have
\eemp
S_1&=\frac{1}{2\pi\xi} \int d\Ttilde{r}\vert\Ttilde{r}\vert^{-1}\widetilde{\rho}(\Ttilde{r}).
\label{s1}
\ffin
Discarding box size effects (assuming $\Delta\to\infty$) $S_1$ behaves as,
\begin{itemize}
\item[$\star$] Mean field (MF - eq. (\ref{rho_inf}))
\eemp
S_{1}^{(MF)}&=\frac{\xi-1-e^{\frac{1}{\xi-1}}\text{E}_1\left[\frac{1}{\xi-1}\right]}{\xi ^2},
\label{S1_MF}
\ffin
\item[$\star$] Strong coupling to leading order (SC-0 - eq. (\ref{rdf}))
\eemp
S_{1}^{(SC-0)}&=f\frac{2(\xi-1)}{\xi(2\xi-1)} =\frac{2(\xi-1)^2}{\xi^2(2\xi-1)} ,
\label{S1_SC-0}
\ffin
\end{itemize}

Figure \ref{fig:Meaninverselength} shows the results from the Monte-Carlo simulations taken with values of $\xi$ and $\Xi$ in all ranges. $S_1$ vanishes for $\xi<1$ since all ions dilute away from the
charged rod, while $S_1 \neq 0$ when $\xi>1$.
As before, the agreement with mean-field is excellent at $\Xi=0.1$,
and equally good is the consistency with SC-0 for large $\Xi$ ($10^3$, $10^4$ and $10^5$). 
The data at $\Xi=10^2$ appear quite close the the strong-coupling limit, but 
exhibit some discrepancy. These data further illustrate the relevance
of normalizing the profile with $f=1-1/\xi$ and not $f=1$, see the upper dotted curve.

\ImT{0.4}{art6_S1_alpha}{Meaninverselength}{Order parameter $S_1$ (mean inverse distance from the cylinder) as a function of the Manning parameter $\xi$, for different couplings $\Xi$ and $\Delta=300$. The  lines indicate the mean field [MF, Eq.~(\ref{S1_MF})] and strong coupling to leading order [SC-0, Eq.~(\ref{S1_SC-0})] analytic solutions. The upper dotted curve corresponds to the analytical prediction, Eq. (68), in Ref. \cite{naji2005counterions,naji1}.}

\subsection{Towards the ground state: crystallization scenario}
\label{ssec:crystallization}

After having investigated the properties of the density profiles perpendicular to the charged rod, we now
address  the question of the correlations along the rod ($z$ direction), through 
the computation of the essentially one dimensional pair correlation function $g_z$ of bounded particles along the $z$-axis. The normalization of this object was done with respect to the number of ions close to the surface of the cylinder, thereby considering in the calculation only
those ions present between two concentric cylindrical shells of at $r=R$ and $r=R+10\mu$.
Normalization ensures that $g(z) \to 1$ at large $z$. 

\psfrag{AX}{$z/a^\prime$}
\psfrag{BY}{$g_z(z)$}
\ImT{0.4}{art8_g_z1}{gz1}{Normalized pair correlation function along the cylinder for
$\xi=3$ and different values of $\Xi$. The number of particles is $N=300$ unless specified in the legend, and $\Delta=300$.}

The results for $g_z$ are presented Fig. \ref{fig:gz1} and \ref{fig:gz2}. 
They shed light on the crystallization phenomenon that takes place here,
and on the asymmetric roles played by the two parameters $\Xi$ and $\xi$.
Fig. \ref{fig:gz1} shows that beyond a certain value, the correlations along
the cylinder no longer depend on $\Xi$. This was expected, from the dual localization 
argument developed in section \ref{ssec:sc1}. A large $\Xi$ confines the ions in the
vicinity of the charged rod, but their interaction along $z$
is governed by $\xi$. If the latter quantity is small, we face
an effective one dimensional liquid, that is mildly modulated in Fig.
\ref{fig:gz1}. In passing, this figure illustrates that the number of
particle taken for the simulations (300) is actually sufficient for our purposes:
identical results are obtained with $N=1000$. 
Upon increasing $\xi$, crystallization occurs along the cylinder, as 
hinted in Fig. \ref{fig:gz2}, where we recover the length scale 
$a'$ as the correct measure of inter-particle distances along $z$.

\ImT{0.4}{art8_g_z_NEW}{gz2}{Pair correlation function along the $z$-axis for different values of $\xi$ and $\Xi=10^4$, and $\Delta=300$. The number of particles is $N=300$ unless specified in the legend. }

From the form of Eq.~(\ref{betaE}) and in particular the harmonic energy term in $z$, 
we expect that the first peak of $g_z$ will present approximate Gaussian shape. 
Under this assumption, the width of the peak can be read directly in Eq. (\ref{betaE}), to be 
\eemp
\delta_z\propto(\xi f)^{-\frac{1}{2}}.
\label{delta_z}
\ffin
From Fig. \ref{fig:gz2}, we can extract $\delta_z$ performing a Gaussian fit of the first peak. 
The resulting width is shown in Fig. \ref{fig:gz_gaussian_fit}, which gives credit to the
naive estimation and show a very good agreement with the linear trend expected for $\delta_z$ as a function of $\delta_z \propto (\xi f)^{-1/2}$.

\psfrag{AX}{$(f\xi)^{-1/2}$}
\ImT{0.4}{art12_delta_z1}{gz_gaussian_fit}{Width of the first peak of the numerical data for 
$g_z$ in fig.~\ref{fig:gz2}. The simple argument giving (\ref{delta_z}) 
leads to expect a straight line. }

\subsection{Ion Profile Close To The Cylinder and correction to SC-0}
\label{ssec:close}

\psfrag{AX}{$r/R$}
\psfrag{BY}{$(r/R)^{2\xi} \,\, \widetilde{\rho}({r})$}
\ImT{0.4}{art7_rho}{Density1}{Radial ionic density for different values of $\xi$, $\Xi=10^2$ and $\Delta=300$. The dotted lines represent the analytic SC-1 result, Eq.~(\ref{rdf_corr}).}

\psfrag{AX}{$r/R$}
\psfrag{BY}{$(r/R)^{2\xi}\,\, \widetilde{\rho}({r})$}
\ImT{0.4}{art9_rho2}{Density2}{Same as Fig. \ref{fig:Density1}, for the same confinement but a higher
coupling parameter $\Xi=10^3$.}

\psfrag{AX}{$r/R$}
\psfrag{BY}{$(r/R)^{2\xi}\, \, \widetilde{\rho}({r})$}
\ImT{0.4}{art10_rho3}{Density3}{Same as Figs. \ref{fig:Density1} and \ref{fig:Density2} for $\Xi=10^4$. The solid curves represent the SC-0* prediction of Eq. (\ref{rdf_numeric}).}

We so far focused on quantities that were accurately described by the SC-0 form
at large $\Xi$. Our goal is now to test the validity of improvements over
this leading form (SC-1 or SC-0*). To this end, we plot in Figs. 
\ref{fig:Density1}, \ref{fig:Density2} and \ref{fig:Density3}
the quantity $r^{2\xi} \widetilde\rho(r)$, that yields a horizontal 
line at SC-0 level, which is a way to make deviations from SC-0 
more apparent. The numerical data thereby obtained show an increase
with $r$, compatible with a parabola,  
which is also the trend predicted by the SC-1 expression, see Eq.
(\ref{rdf_corr}). However,
obtaining a quantitative agreement requires considering large values of 
$\xi$, for the agreement displayed in Fig. \ref{fig:Density1} is quite poor. 
The situation is better in Figs. \ref{fig:Density2} and \ref{fig:Density3}. 
We therefore come to the conclusion that the leading SC-0 behavior
may hold for fairly low values of $\xi$ as discussed in section \ref{ssec:leading},
while upon close inspection, the refinement SC-1 requires $\xi$ to be large.
This comes as no surprise since ionic correlation do not enter the 
SC-0 form, while they are at the root of the SC-1 expression, 
derived assuming that all ions lie in the vicinity of their ground state 
position. As we have seen in section \ref{ssec:crystallization},
this requires typically $\xi>40$, and explains the poor agreement in 
Fig. \ref{fig:Density1}, while we have better consistency 
in Figs. \ref{fig:Density2} and \ref{fig:Density3}. 
In addition, we have reported in Fig. ~\ref{fig:Density3} the results of the alternative 
improvement SC-0* as given by Eq.~(\ref{rdf_numeric}). It seems that such a route 
improves upon SC-0, but also upon SC-1. However, some care is required in 
interpreting the results: while SC-1 follows from an exact although perturbative statistical mechanics
treatment, SC-0* remains at the single particle level, and is heuristic. 
Hence, SC-1 may be viewed as providing the next to leading contribution 
in the SC expansion of the ionic profile, which is not the case of SC-0*.

Note that if $\xi$ is too much increased 
at fixed $\Xi$, the needle requirement $\xi \ll \Xi^{1/2}$ may be
violated at some point. This is not the case though with the data displayed
in Figs. \ref{fig:Density1}, \ref{fig:Density2} and \ref{fig:Density3}.
For instance, we have in the worst case $\xi/\Xi^{0.5} = 0.4$.
It should also be noted here that for large values of the Manning parameter, $(r/R)^{2\xi}$ 
becomes quite large for $r/R>1$; hence, small fluctuations in the density profile induce large fluctuations in the graphed results.

To conclude this section, we report in Fig. \ref{fig:contact} the Monte Carlo measures
for the density profiles at contact, together with the SC-0 and SC-1 predictions. While
SC-0 expectedly gives the correct main trend of $\widetilde \rho(R)$, it is seen
that the $\Xi$ dependent fine structure is well captured by SC-1. 
From the contact theorem, we also know that at fixed $\Xi$, increasing further
$\xi$ ultimately leads to $\widetilde \rho(R)=1$. Such a trend is not visible in 
Fig. \ref{fig:contact}, since the parameter range pertains to the needle limit,
with constraint $\xi \ll \Xi^{1/2}$.  
Additionally, if the limit $\Xi \to \infty$ is taken first at arbitrary $\xi$,
the SC-0 form becomes exact and we have $\widetilde \rho(R) = 2 (\xi-1)^2/\xi^2$,
twice the mean-field expression as already noticed.
Increasing next $\xi$, we finally get the large $\xi$ result 
$\widetilde \rho(R) \to 2$, exactly twice the planar result. It is this
trend that is illustrated in Fig. \ref{fig:contact}. To summarize, the limits
of large $\xi$ and large $\Xi$ do not commute, and we can write
\begin{eqnarray*}
&\displaystyle \lim_{\xi\to\infty} \lim_{\Xi\to\infty} \,\widetilde\rho(R) \,=\, 2  \\
&\displaystyle \lim_{\Xi\to\infty} \lim_{\xi\to\infty} \,\widetilde\rho(R) \,=\, 1. 
\end{eqnarray*}
The latter equality may be written more generally as 
$\lim_{\xi\to\infty} \widetilde\rho(R) \,=\, 1$, for all $\Xi$.

\psfrag{AX}{$\xi$}
\psfrag{BY}{$\widetilde{\rho}(R)$}
\ImT{0.4}{art21_contact}{contact}{Contact density versus Manning parameter. As above, the symbols are for the Monte Carlo data and the lines for the analytical predictions. The dashed line displays the
SC-0 form, which does not depend on $\Xi$. On the other hand, the SC-1 result following from Eq.
(\ref{rdf_corr}) is $\Xi$ dependent, and there are therefore three different branches (continuous curves) 
showing the corresponding expectation for the three values of $\Xi$ studied.}

\section{Finite Size Effects}
\label{sec:finitesize}

\psfrag{AX}{$\log\left(r/R\right)$}
\psfrag{BY}{$\lambda({r})/\lambda$}
\ImT{0.4}{art23_smallDc}{xi3_10Delta10_inflec}{Cumulative density of particles for $\Delta=10$. 
The inflection point is materialized by an arrow.
The value of the condensed fraction, as given by the inflection point criterion, is indicated for each curve.}

In the previous section, we reported results for large system sizes, in order to have 
a universal fraction of condensed ions, and 
a clear cut distinction between those ions that participate in the screening
of the charged rod, and the de-condensed one that lie in the vicinity of the
confining border at $r=D$. These de-condensed ions, which exist for all finite
values of $\xi$, are the precursors of the ions which dissolve away
when the system size is increased to infinity. For smaller systems though,
finite-size effects are very pronounced. This is illustrated in Fig.
\ref{fig:xi3_10Delta10_inflec}, which differs significantly from its large
$\Delta$ counterpart, Fig. \ref{fig:CumulativeDensity}.
Although $\Delta=10$ in the figure is large enough to allow for discriminating
condensed from de-condensed ions, it is seen that the inflection point, which still
provides a convenient cut-off for the partitioning, severely depends on the 
coupling parameter $\Xi$. In the mean-field regime, the corresponding condensed
fraction is still given by $f=f_M=1-\xi^{-1}$ (hence $2/3$ on the figure). 
Beyond mean-field, the quantity increases 
with $\Xi$ in a nontrivial fashion.

A similar conclusion holds concerning the system size dependence,
see Fig. \ref{fig:Plateau2}. For $\Delta>50$ only do we get a condensed fraction 
that is close to its infinite dilution expression, $1-1/\xi = 2/3$ on the figure.
A valid question is then to see if the strong coupling prediction SC-0, Eq. (\ref{rdf}),
holds for the profile,
with proper normalization $f$ following from the inflection point rule. 
We can conclude from Fig. \ref{fig:xi3_10Delta10} that this is indeed
the case: the Monte Carlo data are in good agreement with our prediction,
where the only {\it a priori} unknown is $f$, taken from Fig. 
\ref{fig:xi3_10Delta10_inflec}. The profiles are sandwiched 
between the limiting forms having $f=f_M=1-1/\xi$, which appears to 
be a lower bound for the condensed fraction, and $f=1$.

\psfrag{AX}{$\log\left(r/R\right)$}
\psfrag{BY}{$\lambda({r})/\lambda$}
\ImT{0.4}{art1_charge_Xi100}{Plateau2}{Build-up and size dependence
of the integrated charge plateau for different values of $\Delta$. Here
$\xi=3$ and $\Xi=10^{2}$}

\psfrag{AX}{$r/R$}
\psfrag{BY}{$\widetilde \rho(r)$}
\ImT{0.4}{art23_smallDb}{xi3_10Delta10}{Density profile for $\xi=3$,
$\Delta=10$, against the $f$-normalized strong-coupling prediction. The values of the condensed fraction $f$, used in the SC-0 
form Eq. (\ref{rdf}), are those which are read in Fig. \ref{fig:xi3_10Delta10_inflec}.}{}

\psfrag{AX}{$\Delta$}
\psfrag{BY}{$(f-f_M)/f_M$}
\ImT{0.4}{art4_size1}{size_f1}{Condensed fraction of ions determined numerically using the inflection point criterion in the mean field and weakly coupled regime for $\xi=3$.}

The remaining and final task is to quantify the dependence of $f$ on system size $\Delta$ and coupling 
parameter,
given that $f\simeq f_M$ when $\Delta$ is big enough, see Fig. \ref{fig:Condensedfractionofions}.
To this end, it seems that one should distinguish the regimes of low $\Xi$ where 
one essentially finds the mean-field result $f_M$, see Fig. \ref{fig:size_f1},
from the more strongly correlated cases. Figure 
\ref{fig:size_f2} shows that in the latter case and for fixed coupling $\Xi$, 
$f$ decreases with confinement in such a way that
\emp
\frac{f-f_M}{f_M}\simeq\frac{\varepsilon}{\Delta^\gamma},
\label{size_Delta}
\fin
with $\varepsilon$ and $\gamma$ two dimensionless parameters 
reported in Table \ref{tab:reg_Delta}, and obtained from regressions on all data sets collected.
A first conclusion which can be drawn is that except for too small
$\Delta$ and $\xi$, we have $\gamma \simeq 1$. Quite expectedly,
confining the system favors condensation.
Second, the figure exhibits 
a departure from the $1/\Delta$ scaling on the left hand side, where 
$\Delta$ is small. In this region, the quantity plotted becomes $\Delta$ 
independent, and turns out to reach its maximum possible value,
i.e. $(\xi-1)^{-1}$ corresponding to $f=1$. This is a hint that too small
systems (say below some $\Delta_l$) do not allow for evaporation to set in.
For $\Delta > \Delta_l$, $\Delta (f-f_M) $ becomes $\Delta$ independent,
as also illustrated in Fig. \ref{fig:size_f3}, which furthermore
shows that the $\Xi$ dependence is logarithmic. More precisely, we have
\emp
\frac{f-f_M}{f_M}\simeq\frac{\alpha(\log\Xi-\delta)}{\Delta},
\fin
with $\alpha$ and $\delta$ dimensionless parameters reported in Table \ref{tab:reg_Xi}. All previously 
described effects are encoded within this ultimate expression.
Note however that this result cannot hold for arbitrary large $\Xi$,
for $f$ again has to remain smaller than unity. Hence, 
and similarly to small $\Delta$ results, full condensation with $f=1$ is achieved at large couplings while 
holding a fixed size (see the $\xi=5$ and $\Delta=30$ last two points).

\psfrag{AX}{$\Delta$}
\psfrag{BY}{$(f-f_M)/f_M$}
\ImT{0.4}{art28_size_D}{size_f2}{Condensed fraction under
  strong-coupling, for $\xi=3,4,5$, in a log-log plot where errors are
  less than the tick size. The dashed lines represent linear
  regressions which are summarized in Table \ref{tab:reg_Delta}.}

\begin{table}[!h]
\begin{center}
\begin{tabular}{|c|c||c|c|c|c|}
\hline
$\Xi$ & $\xi$ & $\varepsilon$ & $\sigma_\varepsilon$ & $\gamma$ & $\sigma_\gamma$ \\ \hline\hline
$10^2$ & $3$ & $12.5$ & $2.2$ & $1.99$ & $0.08$ \\ \hline
$10^3$ & $3$ & $7.52$ & $0.06$ & $1.198$ & $0.003$ \\ \hline
$10^4$ & $3$ & $16.0$ & $0.8$ & $1.23$ & $0.02$ \\ \hline\hline
$10^2$ & $4$ & $2.5$ & $0.2$ & $1.57$ & $0.05$ \\ \hline
$10^3$ & $4$ & $3.5$ & $0.2$ & $1.02$ & $0.02$ \\ \hline
$10^4$ & $4$ & $7.6$ & $0.5$ & $1.06$ & $0.02$ \\ \hline\hline
$10^2$ & $5$ & $0.16$ & $0.13$ & $1.0$ & $0.2$ \\ \hline
$10^3$ & $5$ & $4.1$ & $0.1$ & $1.096$ & $0.007$ \\ \hline
$10^4$ & $5$ & $8.4$ & $0.2$ & $1.100$ & $0.005$ \\ \hline\hline
\end{tabular}
\caption{\label{tab:reg_Delta}Numerical values for $\varepsilon$ and $\gamma$ from a regression performed considering 
$\frac{f-f_M}{f_M}\simeq\frac{\varepsilon}{\Delta^\gamma}$ on multiple results. The quantities labeled $\sigma$ refer to the standard 
deviations.}
\end{center}
\end{table}

\psfrag{AX}{$\Xi$}
\psfrag{BY}{$\Delta(f-f_M)/f_M$}
\ImT{0.4}{art29_size_Xi}{size_f3}{Same as Fig. \ref{fig:size_f2}, to
  probe the $\Xi$ dependence in a log-linear plot. The dashed lines
  represent linear regressions that are summarized in Table \ref{tab:reg_Xi}.
  Note that since $f$ is bounded from above by 1 (complete condensation), the quantity
  plotted cannot grow without bounds which explains the departure from scaling
  at large $\Xi$.}

\begin{table}[!h]
\begin{center}
\begin{tabular}{|c|c||c|c|c|c|}
\hline
$\Delta$ & $\xi$ & $\delta$ & $\sigma_\delta$ & $\alpha$ & $\sigma_\alpha$ \\ \hline\hline
$30$ & $4$ & $4.42$ & $0.03$ & $1.37$ & $0.01$ \\ \hline
$50$ & $4$ & $4.45$ & $0.02$ & $1.276$ & $0.007$ \\ \hline
$60$ & $4$ & $4.42$ & $0.03$ & $1.228$ & $0.009$ \\ \hline
$300$ & $4$ & $3.9$ & $0.2$ & $1.01$ & $0.05$ \\ \hline\hline
$30$ & $5$ & $4.50$ & $0.04$ & $1.25$ & $0.01$ \\ \hline
$50$ & $5$ & $4.55$ & $0.03$ & $1.22$ & $0.01$ \\ \hline\hline
\end{tabular}
\caption{\label{tab:reg_Xi}Numerical values for $\alpha$ and $\delta$ from a regression performed considering $\frac{f-f_M}{f_M}\simeq\frac{\alpha(\log\Xi-\delta)}{\Delta}$ on multiple results. The quantities labeled $\sigma$ refer to the standard 
deviations.}
\end{center}
\end{table}

Another interesting feature emerging from Fig. \ref{fig:size_f3} and Table \ref{tab:reg_Xi} is that 
the $x$-axis intercept ($\delta$)
is the same --within numerical accuracy-- for the different sets ($\delta\simeq4.5$). Consequently, for any given value of the coupling 
below $e^\delta$, the system will exhibit ``ideal'' evaporation (Manning evaporation or $f=f_M$) 
regardless of the size and, from the data, also regardless of the Manning parameter. 
This leads us to believe that it is a universal property in the evaporation of ions.
Together with the fact that $\alpha$ takes values close to unity except for too small $\Delta$,
we summarize our finite-size analysis with the expression
\emp
\frac{f-f_M}{f_M}\simeq\frac{(\log\Xi-4.5)}{\Delta}.
\fin
which holds provided $\Xi>e^\delta\simeq 90$ and $\Delta>\Delta_l$, while $f\simeq f_M$ for $\Xi <e^\delta$. Full condensation determines $\Delta_l$ (at $f=1$) in such a way that
\emp
\Delta_l\simeq(\xi-1)(\log\Xi-4.5).
\fin
It should be kept in mind that $\Delta=\log(D/R)$ measures the log-size 
of the system.


\section{Conclusions}
\label{sec:concl}

The behaviour of counter-ions at a planar interface is encoded in the coupling 
parameter $\Xi$, defined from the uniform surface charge $\sigma$ as $\Xi \propto l_B^2 \sigma q^3$.
This scaling simply follows from the fact that the relevant Bjerrum length 
for $q$-valent ions is $q^2 l_B$, and that the natural measure of surface charge 
is $\sigma/q$. Hence the dimensionless charge $(q^2 l_B)^2 \sigma /q$. 
Equivalently, we can view $\Xi$ (or more precisely $\sqrt{\Xi}$) 
as the ratio of thermal energy over the typical 
Coulomb pair repulsion when all ions are condensed onto the plane, and where the
inter-ion distance reads $a \propto \sqrt{q/\sigma}$: $\Xi^{1/2} = (q^2 l_B)/a$.
For large $\Xi$, the ions are confined in a region with extension given by the
Gouy length $\mu \propto (l_B \sigma q)^{-1}$. This applies to all coupling regimes,
from mean-field at small $\Xi$ to strong-coupling \cite{netz2001electrostatistics}. 
When considering curvature,
and addressing cylindrical macro-ions rather than planar, a new parameter enters
the description, $\xi = q l_B\lambda = q l_B 2 \pi R \sigma$. 
Of course, when $R \gg a$, which also means $\xi \gg \Xi^{1/2}$,
the situation is very close to its planar counterpart. It changes significantly
in the opposite case $\xi \ll \Xi^{1/2}$ where curvature is strong, and that we called
here the needle limit. There, the ions are confined in a region of extension 
$R$, given by the radius of the charged rod, that is $\Xi$-independent. In the 
needle regime, the radius $R$ is much smaller than the typical distance
$a'$ between charges along the rod ($R\ll a \ll a'$). This provides 
the rationale for deriving simple strong-coupling (SC) predictions: 
in the sense $R \ll a'$, the ions are far way from each other,
and mostly respond to the log potential of the rod. A single particle picture
holds, that can be seen as the ideal gas behaviour of non interacting
particles in an external field (leaving aside the subtlety of
counter-ion evaporation, that is a collective effect).
Good agreement with Monte Carlo
results can then be achieved, even at ``small'' values of $\xi$. 
This even if, strictly speaking, the ground state of the system is approached
when both $\Xi$ and $\xi$ are large. The reason is that the single particle
picture is already an acceptable approximation when $\xi$ is of order
unity (or slightly beyond), provided $\Xi$ is large. 
We also add that again for large $\Xi$ where the ions remain close to the
cylinder, the Manning parameter $\xi$ controls the essentially one-dimensional
structure of the ionic system, from a repulsive liquid at small $\xi$
to a crystal at large $\xi$. Indeed, although $\xi$ is initially
defined as the dimensionless linear charge of the rod, it can be
rewritten as $q^2 l_B / a'$, and thus qualifies as a one dimensional plasma
parameter, quantifying the strength of interactions for ions along the rod.

An additional feature pertaining to cylindrical charged macromolecules
is the counter-ion condensation-evaporation phenomenon, that plays 
a prominent role here. While its influence on the mean-field behaviour
is well known, we have shown that due account of its effect is essential
for a good agreement between the SC theory and simulations. 
We have explicitly worked out the leading ionic profile under large $\Xi$ 
(SC-0), together with the first correction (SC-1), that indeed improves
upon SC-0, but requires quite large values of $\xi$ to be relevant ($\xi>10$). We also 
illustrated clearly that the ionic profile, even at large $\Xi$, crosses over
to mean-field behaviour far from the plate. This behaviour, 
although expected, is in other settings extremely difficult to 
observe and could only be evidenced due to the (exponentially) large sizes
used in the simulations. 

We have performed a finite-size analysis for the condensed fraction,
which leads to several novel features.  To
this end, we proposed modifications to the previously introduced 
Monte Carlo sampling
method, that significantly improved convergence rate. The signature of 
finite size effects is logarithmic both in the coupling parameter and in system size,
provided $\Xi$ is above $e^\delta\sim90$. 
Full condensation can be achieved for both small
$\Delta$ and large $\Xi$.

Before concluding, we provide some parameter values for an important
rod-like biopolymer. With double-stranded DNA, one has in water at room
temperature ($l_B \simeq 7\,$\AA), one has $\xi\simeq 8$, $\Xi\simeq 22$
with divalent ions $q=2$, and $\xi\simeq 16$, $\Xi\simeq 180$
with tetravalent ions. This latter case is not quite in the
needle limit since $\xi/\Xi^{1/2}$ is of order 1, 
but approaching it. The expressions derived here, which are
salt-free, would then provide a zeroth order limiting case. 

Interesting perspectives opened by this work include dielectric
systems, the study of the effect of salt (added electrolyte),
together with working out the two-dimensional pendant of our 
investigation, where ions interact with a log potential,
which should lead to a large distance physics that is
no longer of mean-field type \cite{Samaj2011counter}.  

We would like to thank Ladislav \v{S}amaj for stimulating discussions
and Martial Mazars for insightful advices on the simulation
code. Support from ECOS Nord/COLCIENCIAS-MEN-ICETEX is
acknowledged. JPM and GT acknowledge partial financial support from
Comité de Investigaciones, Facultad de Ciencias, Universidad de los Andes.

\appendix

\begin{widetext}

\section{Inter-Particle Potential Energy and Density Profile}
\label{U_app}

In this appendix we derive the density profile in the strong coupling
and needle limit up to the first correction in the small ``needle''
parameter $R/a'=\xi^2 f/\Xi$. The following will be the notation used
for normalization: $\widetilde{x}\equiv x/\mu$ and $\check{x}\equiv
x/a^\prime$, and bold letters for vectors accordingly. In this limit,
the distribution of counter-ions presents small thermal fluctuations
from the ground state. Then it is natural to write the position of a
given particle as $\mathbf{R}_l+\mathbf{X}$ with $\mathbf{R}_l=a'l \hat{k}$
and $\mathbf{X}=\mathbf{x}+z \hat{k}$, where $|\mathbf{X}|/|\mathbf{R}_l|$ is of
order $R/a'$. Then,
\eemps
\frac{1}{\vert\mathbf{R}+\mathbf{X}\vert}-\frac{1}{\vert\mathbf{R}\vert}=&-\frac{1}{2}\frac{\mathbf{X}\cdot(2\mathbf{R}+\mathbf{X})}{\vert\mathbf{R}\vert^3}+\frac{3}{2}\frac{(\mathbf{X}\cdot\mathbf{R})^2}{\vert\mathbf{R}\vert^5}+\mathcal{O}((R/a')^3)\\
\approx&\frac{\mu^2}{{a^\prime}^3}\left(-\frac{1}{2}\frac{\Ttilde{X}\cdot(2\Ttilde{R}+\Ttilde{X})}{\vert\Ccheck{R}\vert^3}+\frac{3}{2}\frac{(\Ttilde{X}\cdot\Ccheck{R})^2}{\vert\Ccheck{R}\vert^5}\right),
\ffins
where 
\eemps
\frac{\mu^2}{{a^\prime}^3}&=\frac{1}{\xi^2}\(\frac{R}{{a^\prime}}\)^3\frac{1}{R}
&=\frac{1}{\xi^2}\(\frac{R}{{a^\prime}}\)^3\frac{1}{R}
&=\frac{1}{l_Bq^2}\frac{1}{\xi^3}\(\frac{\xi^2}{\Xi}f\)^3\frac{l_Bq^2\xi}{R}
&=\frac{1}{l_Bq^2}\frac{\xi^3}{\Xi^2}f^3.
\ffins

Therefore, the ion-ion energy term can be written as $\sum_{k<j}
1/(a'|j-k|) + \delta U$, with
\eemps
\beta \delta U=&\frac{1}{2}\frac{\xi^3}{\Xi^2}f^3\sum_{j\neq l}\left\{-\frac{1}{2}\frac{(\Ttilde{X}_j-\Ttilde{X}_l)\cdot(2(\Ttilde{R}_j-\Ttilde{R}_l)+(\Ttilde{X}_j-\Ttilde{X}_l))}{\vert\Ccheck{R}_j-\Ccheck{R}_l\vert^3}+\frac{3}{2}\frac{((\Ttilde{X}_j-\Ttilde{X}_l)\cdot(\Ccheck{R}_j-\Ccheck{R}_l))^2}{\vert\Ccheck{R}_j-\Ccheck{R}_l\vert^5}\right\}\\
=&\frac{\xi^3}{\Xi^2}f^3\sum_{j}\left\{-\frac{1}{2}\sum_{l\neq j}\frac{1}{\vert\Ccheck{R}_j-\Ccheck{R}_l\vert^3}(\widetilde{X}_{j}^{2}+2(\Ttilde{R}_j-\Ttilde{R}_l)\cdot\Ttilde{X}_j)+\frac{3}{2}\sum_{l\neq j}\frac{1}{\vert\Ccheck{R}_j-\Ccheck{R}_l\vert^5}((\Ccheck{R}_j-\Ccheck{R}_l)\cdot\Ttilde{X}_j)^2+\frac{1}{2}\sum_{l\neq j}\frac{1}{\vert\Ccheck{R}_j-\Ccheck{R}_l\vert^3}\Ttilde{X}_j\cdot\Ttilde{X}_l\right.\\
&\left.-\frac{3}{2}\sum_{l\neq j}\frac{1}{\vert\Ccheck{R}_j-\Ccheck{R}_l\vert^5}((\Ccheck{R}_j-\Ccheck{R}_l)\cdot\Ttilde{X}_j)((\Ccheck{R}_j-\Ccheck{R}_l)\cdot\Ttilde{X}_l)\right\}.\\
\ffins
Using $\mathbf{R}_j=a^\prime j\hat{k}$, then $\mathbf{R}_j\cdot\mathbf{X}_l=a^\prime j(z_l)$,
\eemps
\beta\delta U=&\frac{\xi^3}{\Xi^2}f^3\sum_{j}\left\{\frac{-\widetilde{x}_{j}^{2}+2\widetilde{z}_{j}^{2}}{2}\sum_{l\neq j}\frac{1}{\vert j-l\vert^3}+\frac{1}{2}\sum_{l\neq j}\frac{\Ttilde{x}_{l}\cdot\Ttilde{x}_{j}+\widetilde{z}_{l}\widetilde{z}_{j}}{\vert j-l\vert^3}-\frac{3\widetilde{z}_{j}}{2}\sum_{l\neq j}\frac{\widetilde{z}_{l}}{\vert j-l\vert^3}\right\}\\
\ffins
To evaluate the series we use the definition of the Riemann $\zeta$-function ($\zeta(n,q)\colon=\sum_{j=0}^{\infty}\frac{1}{(q+j)^n}$ and $\zeta(n)\colon=\sum_{j=1}^{\infty}\frac{1}{j^n}$).
\eemps
\beta\delta U=&\frac{\xi^3}{\Xi^2}f^3\sum_{j}\left\{\zeta(3)\left(-\widetilde{x}_{j}^{2}+2\widetilde{z}_{j}^{2}\right)+\frac{1}{2}\sum_{l\neq j}\frac{\Ttilde{x}_{l}\cdot\Ttilde{x}_{j}}{\vert j-l\vert^3}-\widetilde{z}_{j}\sum_{l\neq j}\frac{\widetilde{z}_{l}}{\vert j-l\vert^3}\right\}.\\
\ffins

In order to fully analyze the crystallization effect we need to write the energy term separately for perpendicular and parallel displacements to the surface of the cylinder. The strength of the energy cost for displacements parallel to the surface have to be measured against the lattice constant $a^\prime$ (i.e. $\check{z}=z/a^\prime$), thus we have to write $\beta\delta U$ in two separate terms re-normalized appropriately. Hence,

\eemps
\beta\delta U=&\frac{\xi^3}{\Xi^2}f^3\sum_{j}\left\{-\zeta(3)\widetilde{x}_{j}^{2}+\frac{1}{2}\sum_{l\neq j}\frac{\Ttilde{x}_{l}\cdot\Ttilde{x}_{j}}{\vert j-l\vert^3}\right\}+\frac{\xi^3}{\Xi^2}f^3\(\frac{a^\prime}{\mu}\)^2\sum_{j}\left\{2\zeta(3)\check{z}_{j}^{2}-\check{z}_{j}\sum_{l\neq j}\frac{\check{z}_{l}}{\vert j-l\vert^3}\right\},\\
\ffins
that yields the final result for the potential energy change

\eemp
\beta\delta U=&\frac{\xi^3}{\Xi^2}f^3\sum_{j}\left\{-\zeta(3)\widetilde{x}_{j}^{2}+\frac{1}{2}\sum_{l\neq j}\frac{\Ttilde{x}_{l}\cdot\Ttilde{x}_{j}}{\vert j-l\vert^3}\right\}+\xi f\sum_{j}\left\{2\zeta(3)\check{z}_{j}^{2}-\check{z}_{j}\sum_{l\neq j}\frac{\check{z}_{l}}{\vert j-l\vert^3}\right\}.\\
\ffin

Finally, the total energy, up to a constant term, including the
cylinder potential term is as follows
\eemp
\beta\delta E=&\xi\sum_{j}\log\(\frac{x_{j}^{2}}{R^2}\)+\frac{\xi^3}{\Xi^2}f^3\sum_{j}\left\{-\zeta(3)\widetilde{x}_{j}^{2}+\frac{1}{2}\sum_{l\neq j}\frac{\Ttilde{x}_{l}\cdot\Ttilde{x}_{j}}{\vert j-l\vert^3}\right\}+\xi f\sum_{j}\left\{2\zeta(3)\check{z}_{j}^{2}-\check{z}_{j}\sum_{l\neq j}\frac{\check{z}_{l}}{\vert j-l\vert^3}\right\}.\\
\ffin

The particle density profile can be calculated as $\rho(\mathbf{x})=C\langle\delta(\widetilde{x}-\widetilde{x}_0)\rangle$ (with $C$ a normalizing constant satisfying eq. \ref{new_norm_cond}). Notice that we have chosen the perpendicular displacements for the density profile.
\eemp
\widetilde{\rho}(\Ttilde{x})=&C\(\prod_{k\neq0}\int
d\Ttilde{x}_k\)\(\prod_l
\int dz_l\)\exp\[-\beta\delta E\],
\ffin
where integral over the perpendicular displacements $\Ttilde{x}_k$ is 
over $N-1$ particles. Expanding $\exp(-\beta \delta U)$ up to the first
correction in $R/a'$, we have 
\eemps
\widetilde{\rho}(\Ttilde{X})=&C_0\(\frac{R^2}{x^{2}}\)^\xi\(\prod_{k\neq0}D_k\int
d\Ttilde{x}_k\(\frac{R^2}{x_{k}^{2}}\)^\xi\)\left\{1+\frac{\xi^3}{\Xi^2}f^3\left\{\zeta(3)\left(x^2+\sum_{j\neq
  0}\widetilde{x_j}^{2}\right)-\sum_{j<l}\frac{\Ttilde{x}_{j}\cdot\Ttilde{x}_l}{\vert j-l\vert^3}\right\}+o((R/a')^2)\right\},
\ffins
with $C_0$ and $D_k$ normalization constants. The contributions from
the $z$ direction has been integrated out. We define $\Omega(R,\Ttilde{x})$ a distribution as
\emp
\Omega(\widetilde{R},\Ttilde{x})\colon=D\(\frac{\widetilde{R}^2}{\widetilde{x}^{2}}\)^\xi,
\fin
where $D=\frac{\xi-1}{\pi\widetilde{R}^2}\,,\forall\xi\in\Re\wedge\xi>1$ is a constant that normalizes the distribution. The domain of $\Omega$ for $x$ is $x\in\left[R,\infty\right),\ \theta\in\left[0,2\pi\right]$. To perform the calculations we will use the simplified notation
\emp
\langle f\rangle_k\colon=\int d\Ttilde{x}_k\Omega(\widetilde{R},\Ttilde{x}_k)f,
\fin
where $\langle1\rangle_k=1$. Using the latter notation, we notice that
the probability distributions for different particles are uncorrelated 
$\langle\langle\Ttilde{x}_j\cdot \Ttilde{x}_k
\rangle_k\rangle_j=\langle\Ttilde{x}_j\rangle_j\cdot
\langle\Ttilde{x}_k\rangle_k$, for $k\neq j$, and due to the
cylindrical symmetry  $\langle\Ttilde{x}_j\rangle_j=0$. Furthermore
$\langle{\Ttilde{x}_k}^2\rangle_k=\xi^2(\xi-1)/(\xi-2)$. Then,
the density is
\eemp
\widetilde{\rho}(x)=&C_0\(\frac{R}{x}\)^{2\xi}\left(1+\zeta(3)\frac{\xi^3}{\Xi^2}f^3\left(\widetilde{x}^{2}+(N-1)\xi^2\frac{\xi-1}{\xi-2}\right)+o((R/a')^2)\right).
\ffin
We proceed to evaluate $C_0$ such that $\int d\Ttilde{r}\,\widetilde{\rho}(\Ttilde{r})=2\pi f\xi$,
\eemps
C_0=\frac{2f \xi(\xi-1)}{\xi^{2(1-\xi)}}
\left(1-\zeta(3)\frac{\xi^3}{\Xi^2}f^3 N \xi^2\frac{\xi-1}{\xi-2}
+o((R/a')^2)\right)
\ffins
where $f$ is the ratio between the condensed integrated charge and the
cylinder surface charge. The density profile is
\eemp
\widetilde{\rho}_{SC}(\widetilde{r})=&{f}\frac{2(\xi-1)}{\xi}\(\frac{R}{r}\)^{2\xi}\left(1+\zeta(3)\frac{\xi^5}{\Xi^2}{f}^3\left(\left(\frac{r}{R}\right)^{2}-\frac{\xi-1}{\xi-2}\right)
+o((R/a')^2)
\right),\\
\ffin
which yields a value at contact equal to
\eemp
\widetilde{\rho}_{SC}(R)=&{f}\frac{2(\xi-1)}{\xi}\left(1-\frac{\zeta(3)}{\xi-2}\frac{\xi^5}{\Xi^2}{f}^3
+o((R/a')^2)\right).\\
\ffin

\section{About the 1D Ewald Summation}
\label{1D-Ewald}

Let us determine the Ewald summation for a one-dimensional periodic system. First, we write the potential as a summation of one particle interaction and its images at a point $\mathbf{r}$ ($\mathbf{L}_n=n L\hat{k}$ for $n\in\mathbb{Z}$)
\eemp
\phi&=\sum_{\mathbf{L}_n}\frac{1}{\vert\mathbf{r}+\mathbf{L}_n\vert}\\
&=\sum_{\mathbf{L}_n}\left[\frac{1}{\sqrt{\pi}}\int_{0}^{\infty}\frac{dt}{\sqrt{t}}e^{-\vert\mathbf{r}+\mathbf{L}_n\vert^2t}\right]\\
&=\sum_{\mathbf{L}_n}\left[\frac{1}{\sqrt{\pi}}\int_{\alpha^2}^{\infty}\frac{dt}{\sqrt{t}}e^{-\vert\mathbf{r}+\mathbf{L}_n\vert^2t}\right]+\sum_{\mathbf{L}_n}\left[\frac{1}{\sqrt{\pi}}\int_{0}^{\alpha^2}\frac{dt}{\sqrt{t}}e^{-\vert\mathbf{r}+\mathbf{L}_n\vert^2t}\right],\\
\ffin
with $\alpha$ a real positive parameter chosen wisely for convergence. The first integral is the well known $\text{Erfc}(x)/x$ function. Then using the Poisson-Jacobi transformation for the second sum (exchanging sum and integral)
\eemp
\phi=&\sum_{\mathbf{L}_n}\frac{\text{Erfc}(\alpha\vert\mathbf{r}+\mathbf{L}_n\vert)}{\vert\mathbf{r}+\mathbf{L}_n\vert}+\frac{1}{\sqrt{\pi}}\int_{0}^{\alpha^2}\frac{dt}{\sqrt{t}}\left[\sum_{\mathbf{L}_n}e^{-\vert\mathbf{r}+\mathbf{L}_n\vert^2t}\right]\\
=&\sum_{\mathbf{L}_n}\frac{\text{Erfc}(\alpha\vert\mathbf{r}+\mathbf{L}_n\vert)}{\vert\mathbf{r}+\mathbf{L}_n\vert}+\frac{1}{\sqrt{\pi}}\int_{0}^{\alpha^2}\frac{dt}{\sqrt{t}}\left[e^{-\rho^2t}\sum_{\mathbf{L}_n}e^{-(z+nL)^2t}\right]\\
=&\sum_{\mathbf{L}_n}\frac{\text{Erfc}(\alpha\vert\mathbf{r}+\mathbf{L}_n\vert)}{\vert\mathbf{r}+\mathbf{L}_n\vert}\\
&+\frac{1}{\sqrt{\pi}}\int_{0}^{\alpha^2}dt\frac{e^{-\rho^2t}}{\sqrt{t}}\left[\frac{1}{L}\sum_{k}\left(\sqrt{\frac{\pi}{t}}e^{-\frac{k^2}{4t}}e^{ikz}\right)\right],\\
\ffin
where $k=\frac{2\pi}{L}n$ for $n\in\mathbb{Z}$. Simplifying and considering independently the term $k=0$ from the rest,
\eemp
\phi&=\sum_{\mathbf{L}_n}\frac{\text{Erfc}(\alpha\vert\mathbf{r}+\mathbf{L}_n\vert)}{\vert\mathbf{r}+\mathbf{L}_n\vert}+\frac{1}{L}\sum_{k\neq0}e^{ikz}\left[\int_{0}^{\alpha^2}dt\frac{e^{-\rho^2t-\frac{k^2}{4t}}}{t}\right]+\frac{1}{L}\left[\int_{0}^{\alpha^2}dt\frac{e^{-\rho^2t}}{t}\right].\\
\ffin
The second integral can be rewritten by substituting $u=\frac{\alpha^2}{t}$ ($\frac{du}{u}=-\frac{dt}{t}$)
\eemp
\phi&=\sum_{\mathbf{L}_n}\frac{\text{Erfc}(\alpha\vert\mathbf{r}+\mathbf{L}_n\vert)}{\vert\mathbf{r}+\mathbf{L}_n\vert}+\frac{1}{L}\sum_{k\neq0}e^{ikz}\left[\int_{1}^{\infty}du\frac{e^{-\frac{k^2}{4\alpha^2}u-\frac{\alpha^2\rho^2}{t}}}{u}\right]+\frac{1}{L}\left[\int_{0}^{\alpha^2}dt\frac{e^{-\rho^2t}}{t}\right],\\
\ffin
where the second integral is by definition~(\ref{inc-Bessel}) the incomplete Bessel function $\mathbf{K}_0(\frac{k^2}{4\alpha^2},\alpha^2\rho^2)$, which can be represented with the uniformly convergent series~\cite{Harris2008,Harris2009}
\emp
\mathbf{K}_0(x,y)=\Gamma(0,x+y)J_0(2y)+\sum_{n=1}^{\infty}\left[\frac{\Gamma(n,x+y)}{(x+y)^n}+(-1)^n(x+y)^n\Gamma(-n,x+y)\right]
J_n(2y),
\fin
with $\Gamma(n,x)$ and $J_n(x)$ the incomplete Gamma Function and the Bessel function respectively. The evaluation of the function for the $y>x$ takes some time but we can use the following identity,
\emp
\mathbf{K}_0(x,y)=2K_0(2\sqrt{xy})-\mathbf{K}_0(y,x),
\fin
with $K_0$ the modified Bessel function of order 0. 
Therefore, we can write the overall expression
\eemp
\phi&=\sum_{\mathbf{L}_n}\frac{\text{Erfc}(\alpha\vert\mathbf{r}+\mathbf{L}_n\vert)}{\vert\mathbf{r}+\mathbf{L}_n\vert}+\frac{1}{L}\sum_{k\neq0}e^{ikz}\mathbf{K}_0\left(\frac{k^2}{4\alpha^2},\alpha^2\rho^2\right)+\frac{1}{L}\left[\int_{0}^{\alpha^2}dt\frac{e^{-\rho^2t}}{t}\right].\\
\ffin

Taking $\epsilon>0$,
\eemp
\int_{0}^{\alpha^2}dt\frac{e^{-\rho^2t}}{t}&=\lim_{\epsilon\to0^+}\int_{\epsilon}^{\alpha^2}dt\frac{e^{-\rho^2t}}{t}\\
&=\lim_{\epsilon\to0^+}\int_{\epsilon}^{\infty}dt\frac{e^{-\rho^2t}}{t}-\int_{\alpha^2}^{\infty}dt\frac{e^{-\rho^2t}}{t}\\
&=\lim_{\epsilon\to0^+}\text{E}_1(\epsilon\rho^2)-\text{E}_1(\alpha^2\rho^2)\\
&=-\gamma-\log(\alpha^2\rho^2)-\text{E}_1(\alpha^2\rho^2)+\log\alpha^2-\lim_{\epsilon\to0^+}\log\epsilon,\\
\ffin
where $\text{E}_1(x)$ is the exponential integral. Substituting,
\eemp
\phi&=\sum_{\mathbf{L}_n}\frac{\text{Erfc}(\alpha\vert\mathbf{r}+\mathbf{L}_n\vert)}{\vert\mathbf{r}+\mathbf{L}_n\vert}+\frac{1}{L}\sum_{k\neq0}e^{ikz}\mathbf{K}_0\left(\frac{k^2}{4\alpha^2},\alpha^2\rho^2\right)+\frac{1}{L}\left[-\gamma-\log(\alpha^2\rho^2)-\text{E}_1(\alpha^2\rho^2)+\log\alpha^2-\lim_{\epsilon\to0^+}\log\epsilon\right].\\
\ffin

The self energy term is found following the latter procedure excluding from the summation the $\mathbf{L}=0$ term,
\eemp
\phi_0&=\sum_{\mathbf{L}_n\neq0}\frac{\text{Erfc}(\alpha\vert\mathbf{L}_n\vert)}{\vert\mathbf{L}_n\vert}+\frac{1}{L}\sum_{k\neq0} \mathbf{K}_0\left(\frac{k^2}{4\alpha^2},0\right)+\frac{1}{L}\left[\log\alpha^2-\lim_{\epsilon\to0^+}\log\epsilon\right]-\frac{1}{\sqrt{\pi}}\int_{0}^{\alpha^2}\frac{dt}{\sqrt{t}}\\
&=\sum_{\mathbf{L}_n\neq0}\frac{\text{Erfc}(\alpha\vert\mathbf{L}_n\vert)}{\vert\mathbf{L}_n\vert}+\frac{1}{L}\sum_{k\neq0}\text{E}_1\left(\frac{k^2}{4\alpha^2}\right)+\frac{1}{L}\left[\log\alpha^2-\lim_{\epsilon\to0^+}\log\epsilon\right]-\frac{2\alpha}{\sqrt{\pi}}\\
\ffin
Finally, we consider the contribution due to the cylinder within the cell and the particles and with its images. The following is the result of integrating with respect to $z$ which leads the following two results,
\eemp
\phi_{cyl}&=\text{E}_1(\alpha^2\rho^2)+\left[-\gamma-\log(\alpha^2\rho^2)-\text{E}_1(\alpha^2\rho^2)+\log\alpha^2-\lim_{\epsilon\to0^+}\log\epsilon\right]\\
&=-\gamma-\log(\alpha^2\rho^2)+\log\alpha^2-\lim_{\epsilon\to0^+}\log\epsilon,\\
\ffin
\eemp
\phi_{cyl0}&=L\sum_{\mathbf{L}_n\neq0}\frac{\text{Erfc}(\alpha\vert\mathbf{L}_n\vert)}{\vert\mathbf{L}_n\vert}+\sum_{k\neq0}\text{E}_1\left(\frac{k^2}{4\alpha^2}\right)+\left[\log\alpha^2-\lim_{\epsilon\to0^+}\log\epsilon\right]-\frac{2\alpha L}{\sqrt{\pi}}.\\
\ffin
The overall energy $U$ can be written as
\eemp
4\pi\epsilon\epsilon_0 U=\frac{1}{2}\sum_{i\neq j}q_iq_j\phi(\mathbf{r}_{ij})+\frac{1}{2}\sum_{i}q_{i}^{2}\phi_0+\frac{1}{2}L\lambda^2\phi_{cyl0}+\lambda\sum_{i}q_i\phi_{cyl}(\mathbf{r_i}),
\ffin
with $\lambda$ the linear charge density of the cylinder ($\sigma=\frac{\lambda}{2\pi R}$). Notice that the cylinder is located at the origin of the coordinates. For neutral systems ($\sum_{i}q_i+L\lambda=0$),
\eemp
U=U_{R}+U_{R}+U_{C}+U_{S},
\ffin
with,
\eemp
U_{R}=&\frac{1}{4\pi\epsilon\epsilon_0}\sum_{i=1}^{N-1}\sum_{j=i+1}^{N}q_iq_j\left[\sum_{n}\frac{\text{Erfc}\(\alpha\(\rho_{ij}^{2}+(z_{ij}+Ln)^2\)^{\frac{1}{2}}\)}{\(\rho_{ij}^{2}+(z_{ij}+Ln)^2\)^{\frac{1}{2}}}+\frac{1}{L}\begin{cases}
0,&\mbox{if } \rho_{ij}=0 \\
-\gamma-\log\(\alpha^2\rho_{ij}^{2}\)-\text{E}_1\(\alpha^2\rho_{ij}^{2}\),&\mbox{otherwise} \\
\end{cases}\right],
\ffin
\normalsize
\eemp
U_{F}=\frac{2}{L}\frac{1}{4\pi\epsilon\epsilon_0}\sum_{i=1}^{N-1}\sum_{j=i+1}^{N}q_iq_j\sum_{k>0}\mathbf{K}_{0}\(\frac{k^2}{4\alpha^2},\alpha^2\rho_{ij}^{2}\)\cos(k\cdot z_{ij}),
\ffin
\eemp
U_{C}=&-2\frac{1}{4\pi\epsilon\epsilon_0}\lambda\sum_{i=1}^{N}q_i\log\(\frac{\rho_{i}}{R}\),
\ffin
\eemp
U_{S}=&\frac{1}{4\pi\epsilon\epsilon_0}\(\sum_{n>0}\frac{\text{Erfc}(\alpha Ln)}{Ln}+\frac{1}{L}\sum_{k>0}\text{E}_1\(\frac{k^2}{4\alpha^2}\)-\frac{\alpha}{\sqrt{\pi}}\)\(\sum_{i=1}^{N}q_{i}^{2}+\lambda^2L^2\)+\frac{1}{4\pi\epsilon\epsilon_0}\lambda^2L(\gamma+\log(\alpha^2R^2)).
\ffin
Note that the term $\left[\log\alpha^2-\lim_{\epsilon\to0^+}\log\epsilon\right]$ in the overall energy cancels due to electro-neutrality. This final result is free of divergences and is absolutely convergent for all ranges of the variables $\rho_{i}$, $\rho_{ij}$ and $z_{ij}$.

\end{widetext}

\section{Special Functions}
\label{SF}

The following are definitions used throughout the paper.

\begin{itemize}
\item[$\ddag$] The complementary error function $\text{Erfc}(x)$
\eemp
\text{Erfc}(x)=&\frac{2}{\sqrt{\pi}}\int_{x}^{\infty}e^{-t^2}dt\\
=&\frac{2x}{\sqrt{\pi}}\int_{1}^{\infty}e^{-t^2x^2}dt.
\ffin
\item[$\ddag$] The incomplete gamma function $\Gamma(x,y)$,
\eemp
\Gamma(x,y)=\int_{y}^{\infty}t^{x-1}\exp(-t)dt.
\ffin
\item[$\ddag$] The exponential integral $\text{E}_1(x)$
\eemp
\text{E}_1(x)=&-\gamma-\log x+\sum_{k=1}^{\infty}\frac{(-1)^{k+1}x^k}{k k!}\\
=&\int_{1}^{\infty}\frac{e^{-tx}}{t}dt.
\ffin
Also,
\eemp
\frac{d}{dx}\text{E}_1(x)=-\frac{e^{-x}}{x},\\
\ffin
and the 2D Fourier transform is
\eemp
\mathcal{F}[\text{E}_1(\alpha^2r^2)]({\mathbf{k}})=4\pi\frac{1-e^{-\frac{{k}^2}{4\alpha^2}}}{{k}^2}.
\ffin
\item[$\ddag$] The incomplete Bessel function $\mathbf{K}_{\nu}(x,y)$
\begin{equation}
\label{inc-Bessel}
\mathbf{K}_{\nu}(x,y)=\int_{1}^{+\infty}\frac{e^{-xt-\frac{y}{t}}}{t^{\nu+1}}dt\,.
\end{equation}
\end{itemize}

\vspace{0.5cm}

\bibliography{bibliography}

\end{document}